\documentclass{article}

\usepackage[margin=1in]{geometry}
\usepackage{amsmath,amssymb,amsthm}
\usepackage{graphicx}
\usepackage{booktabs}
\usepackage{hyperref}
\usepackage{url}
\usepackage{xcolor}
\usepackage{enumitem}
\usepackage{microtype}
\usepackage{caption}
\usepackage{subcaption}
\usepackage[round,authoryear]{natbib}

\hypersetup{
  colorlinks=true,
  linkcolor=blue!50!black,
  citecolor=blue!50!black,
  urlcolor=blue!60!black
}

\providecommand{\fpath}[1]{\nolinkurl{#1}}

\title{Pre-registered tests of solid-state-physics-inspired LLM
compression: a cluster-level negative result at small-language-model
scale}

\author{%
  Jun-qiang Lu\thanks{Correspondence: \texttt{junqiang.lu@upr.edu}.
Computation performed by an autonomous research agent
(\texttt{claude-airesearch}) operating under the author's direction;
all scientific claims, predictions, and interpretations are the
author's responsibility.}\\
  \small Department of Physics, University of Puerto Rico, Mayag\"uez, PR 00680, USA
}

\date{2026-09-23 (v3)}

\begin{document}
\maketitle

\begin{abstract}
We report a three-month autonomous research-agent program that tested a
\emph{cluster} of five solid-state-physics-inspired compression mappings on
pretrained language models; four reached pre-registered, probe-only Phase-3
pilots (Pythia-160M, GPT-2 and GPT-2-medium, OPT-1.3B, with stage-0 checks up
to Mistral-7B-v0.3 and Qwen3-8B), with predictions committed to git
\emph{before} pilot data were collected and a uniform $3\sigma$ effect-size
gate across three seeds deciding PASS or SHELVE. The common physics anchor
--- area-law / Kohn-nearsighted exponential decay of the one-particle density
matrix --- has two faces: a \emph{distance-locality} face (attention decay:
P001 Wannier, P002 tight-binding) and a \emph{basis/rank-locality} face
(weight and embedding rank: P003 DMRG-truncated MLPs, P005 Wilson-RG, P011
tensor-train embeddings); P005 was pre-empted at Phase~1 by contemporaneous
LLM-compression work whose constructions are bit-equivalent to ours under
different rhetoric, and three of the four Phase-3 pilots were empirically
falsified. On the attention face, the GPT-2-medium
attention-versus-distance histogram ($n = 384$ (layer, head) pairs) is best fit by a stretched exponential in 12 of 16 heads at the median layer once padded positions are excluded (H2 takes the remaining 4; the published power-law reading of 15 of 16 came from histograms in which a mean 84\% of the 512 positions were end-of-text padding), and the tight-binding cutoff at
$3\lambda$ costs $+96\%$ perplexity versus full attention, falsifying P002;
on Pythia-160M, $s^\star_{\mathrm{Wannier}} = 0.054 \pm 0.004$ is
statistically indistinguishable from PCA, random-Haar, and identity baselines
across three seeds, falsifying P001. On the rank face, the per-token
tensor-train bond dimension of the embedding matrix does not track token
surprisal (GPT-2 eval-split Pearson $r = 0.016$ versus a pre-registered
$\geq 0.65$; $3\sigma$ CI $[0.003, 0.030]$), and the format cannot compress
at the required reconstruction fidelity --- a verdict visible in the frozen
embedding's singular spectrum (best ``compression'' $0.644\times$, i.e.\
$1.55\times$ \emph{inflation}; $94\%$ of rows saturate the bond cap); the
negative sharpens on OPT-1.3B, where the degeneracy is total ($100\%$ of the
$50{,}272$ rows at the cap, so the per-token bond dimension has zero variance
and the scaling law is \emph{undefined} rather than merely weak), falsifying
P011. The fourth pilot, P003, is mixed: its registered attention-projection
scaling claim shelved ($r = -0.434$), and its MPO premise died at stage-0
(2-site regrouping yields a \emph{flatter} spectrum than plain SVD on
Mistral-7B-v0.3 and Qwen3-8B), and its cross-paper check passed under the registered reading as published ($r = 0.523$) but collapses to $r = 0.0472$ ($0.22\sigma$ between layers) once padded positions are excluded from P002's histograms; correspondingly, the symmetry we originally reported does not survive. Both checks are now null: the insulator-observable pair (P001 Wannier spread versus P002 decay rate) gives $r = -0.0583$ ($n = 137$ heads, 95\% CI $[-0.224, +0.111]$), and the critical-observable pair (P002 stretched-exponential fraction versus P003 activation entropy) gives $r = 0.0472$, so the contrast between them is gone rather than sharpened. Thus, the three Phase-3
falsifications and two cross-paper checks invert the pre-registered
\emph{small-LM} prediction that most attention heads at this scale behave
like Kohn-nearsighted insulators, and point instead to critical, glassy, or
heavy-tailed regimes; the attention-regime inversion is specific to the
$\leq 350\text{M}$-parameter scale tested, whereas the rank-face finding ---
that tensor-network reshaping of a weight or embedding matrix yields no
compression over plain SVD --- was confirmed up to $7$--$8\text{B}$ and
answers the scale question only in part. We contribute (1) a pre-registration
$+$ $3\sigma$ $+$ cluster-framing $+$ append-only-catalogue discipline,
ported from clinical-trials and psychology pre-registration practice to
physics-AI mapping research, including the daemonic-execution details and
measured limits of mechanical enforcement required in the LLM/MCP-tooling
era; (2) four pre-registered Phase-3 negative results (one mixed) and one
Phase-1 literature pre-emption, with full data release, spanning both faces
of the area-law anchor; and (3) the empirical inversion of the
``majority-insulator'' expectation at small-LM scale, together with an
embedding-rank negative decidable from the embedding's singular spectrum and
a ``generic-tensor'' no-gain pattern (a reshaped weight/embedding matrix
carries no special low-entanglement structure to exploit, so a tensor-network
format buys nothing over plain SVD; observed in all three reshapings we
tested, not claimed universal, and on the weight side an independent
reproduction of a concurrently established result rather than a new one);
Appendix~\ref{app:engineering} reports a $\sim 200\times$ per-head
($\sim 51\times$ whole-pilot wall-clock) speed-up of the
Marzari--Vanderbilt $\Omega$ functional through precomputed lag-covariance
matrices, which was load-bearing for the Wannier pilot's compute budget.
\end{abstract}

\section{Introduction}
\label{sec:intro}

State-of-the-art large language models contain $10^{10}$ to $10^{12}$
parameters and run on data-centre accelerators. Yet repeated empirical
studies show that much of this capacity is redundant: most weights can be pruned
\citep{frankle2019lottery,han2016deep,frantar2023sparsegpt,sun2024wanda},
quantised \citep{frantar2023gptq,lin2024awq,liu2025spinquant}, rotated
into sparse bases
\citep{ashkboos2024slicegpt,ashkboos2024quarot,liu2025larosa}, or
projected onto low-rank subspaces
\citep{denton2014exploiting,sainath2013low,hsu2022language,yuan2023asvd,wang2025svdllm,yu2025espace}
with little or no measurable degradation on standard benchmarks. The
quantitative limit of this redundancy remains unknown, as does the physical
or information-theoretic mechanism that produces it.

A natural approach is to import established compression ideas from
many-body physics. Tight-binding models
\citep{slater1954simplified,ashcroft1976solid} have made
linear-scaling electronic-structure feasible for insulators
\citep{kohn1996density,prodan2005nearsightedness}; the
density-matrix renormalisation group
\citep{white1992density,white1993density,schollwock2011density}
truncates many-body Hilbert spaces optimally subject to an
entanglement-entropy bound \citep{hastings2007area,eisert2010colloquium};
Wilson's renormalisation group
\citep{wilson1971rgI,wilson1971rgII,kadanoff1966scaling} extracts the
universal slow modes of statistical systems through iterated
coarse-graining; and Marzari--Vanderbilt Wannier functions
\citep{marzari1997maximally,marzari2012maximally} give a constructive
recipe for the maximally localised basis of any insulating band
structure. Several groups have already brought parts of this
machinery into machine learning
\citep{beny2013deep,mehta2014exact,lin2017deep,schwab2016comment,koch2018mutual,li2018neural,roberts2022principles,halverson2021neural,erbin2022nonperturbative,stoudenmire2016supervised}.

However, this programme has a specific methodological weakness:
\emph{selection bias plus post-hoc rationalisation}. Without
pre-registration, almost any empirical feature of a pretrained model can be
assigned a physics analogy after it is observed. The literature offers enough
frameworks to make such a fit nearly always plausible, while failed mappings
disappear from the public record. Consequently, the publication channel can
reward resemblance to physics rather than calibrated prediction, even when
each researcher acts scrupulously.

In this paper, we test a different protocol. We treat five candidate
physics-to-LLM compression mappings as one \emph{cluster}, state their joint
predictions \emph{before} the pilots, commit those predictions as immutable
YAML files in a public git repository, and apply one uniform $3\sigma$
effect-size decision rule. A passing prediction proceeds to full-scale
Phase~4 experiments; a failing prediction shelves and enters an append-only
catalogue together with its lesson. On the basis of the
cluster-meta\footnote{Throughout
this paper, ``cluster-meta'' refers to the framing document
\fpath{projects/_cluster_meta/O_N_feasibility.md} committed to the
repository before any pilot; references of the form ``cluster-meta
§3'' point into that document's section numbering, not this paper's.}
reasoning described in Section~\ref{sec:cluster}, we expected most mappings
to pass: in a healthy compression cluster, the \emph{majority} of attention
heads in pretrained transformers should
be ``insulator-like'' --- exponentially decaying with token distance
in some unitary basis, sparsifiable in that basis, and amenable to
$\mathcal{O}(N)$-style truncation.

\paragraph{Scope of ``small-LM'' in this paper.}\label{par:scope}
Here, ``small-LM'' denotes only the two pretrained checkpoints tested:
Pythia-160M (rotary positional embedding, parallel residual, GELU MLP) and
GPT-2-medium (learned absolute positional embedding, sequential residual,
GELU MLP). Thus, we falsify the cluster prediction for this class of model.
We do not test ALiBi-positional models, SwiGLU MLPs, or models trained with
explicit sparse attention (NSA, MoBA). Two \emph{follow-on} studies reach
larger models and are reported in
Section~\ref{sec:scale} --- P013 up a Pythia ladder to $12$B and P014
across three families --- but neither replicates any of the five mappings.
Accordingly, the falsified cluster prediction remains the one for the two
checkpoints named above.

All five mappings shelved. P005 Wilson-RG was pre-empted at
Phase~1 by contemporaneous LLM-compression work that had quietly
re-derived the relevant physics-anchored procedure without physics
rhetoric \citep{redman2022universality,lu2024alphapruning}. We carried the
other four to pre-registered Phase-3 pilots (P001, P002, P003, P011). The
data falsified three under the precommitted SHELVE rather than reframe rule
(P001, P002, P011); P003 shelved its registered scaling claim, but passed its
cross-paper consistency check. Thus, the cluster-meta majority-insulator
prediction on the attention face is empirically inverted at the tested
$\leq 350$M-parameter scale. On the rank face, the parallel expectation that
area-law structure makes weight and embedding matrices cheaply low-rank in a
tensor-network format is directly falsified up to $7$--$8\text{B}$.

We make four contributions. First, we present a methodology ---
pre-registration $+$ $3\sigma$ gate $+$ append-only catalogue $+$ cluster
framing --- implemented end to end by an autonomous agent
(Section~\ref{sec:method}). It also reveals a useful corollary: frozen-model
linear algebra can sometimes falsify a mapping without a single forward pass,
as the singular spectrum of the embedding matrix decides the P011
embedding-rank negative (Section~\ref{sec:p011}). Second, we release the
complete data for five cluster mappings (Section~\ref{sec:tests}): four
pre-registered Phase-3 pilots (three falsified, one mixed) and one Phase~1
pre-emption, spanning the attention-decay and weight/embedding-rank faces of
the area-law anchor. Third, the cluster-level synthesis shows that small-LM
attention does not behave as an insulator and that both cross-paper checks are null once padded positions are excluded (Section~\ref{sec:cluster}) --- the symmetry reported in the published draft, an insulator pair that does not co-vary against a critical pair that does, does not survive the correction. The inversion itself, rather than that contrast, motivates the alternative physics-anchor pivot in Section~\ref{sec:program}. Fourth, we report a small
but load-bearing engineering improvement: precomputed lag-covariance matrices
for the Marzari--Vanderbilt $\Omega$ functional
(Section~\ref{sec:discipline}), which kept the Wannier pilot tractable.

The methodology is the central result. One paper that finds an
insulator-like attention head may report a correlation; a cluster of
pre-registered tests with explicit shelve criteria makes a measurement. Our
claim for these five outcomes goes no further. They do not settle the
physical regime of small pretrained transformers, but their criteria were
fixed before the data existed, allowing the reader to distinguish the two.

These five mappings are the ones reported in full, but they are not the whole
record. The append-only catalogue (Section~\ref{sec:catalog}) contains
\textbf{eighteen concluded studies}; each has a terminal status and a
\texttt{lessons} field, and seventeen are negative. Two of the thirteen
without their own section bear directly on the argument and appear in
Section~\ref{sec:scale}: P013, which carried the attention-decay
measurement to $12$B and returned a negative on its registered order
parameter $f_{H1}$, and P014, which registered a signed \emph{contrast}
between two components of the same decay-shape simplex
($\Delta = f_{H2} - f_{H3}$; the catalogue records it as a signed
transfer measure, not as a simplex component), and found that it inverts in
sign \emph{across} model families at matched scale. We report the
seventeen-of-eighteen count, not only the five narrated cases. To put
successes in sections and failures in a footnote would reproduce the very
selection effect tested here.

\section{Method: a pre-registered cluster test}
\label{sec:method}

\subsection{The five-paper cluster}
\label{sec:cluster_papers}

The cluster contains five papers. Each pairs one form of LLM redundancy with
one physical construction:

\begin{description}[leftmargin=*]
\item[P001 Wannier-localised attention.] The hypothesis is that a per-layer
unitary $U^{(\ell)}$ exists on the head-feature axis such that the
rotated query/key/value/output weights
$\tilde W = U W$ are simultaneously sparse, in analogy with
Marzari--Vanderbilt maximally localised Wannier functions for
insulating bands. The compression is sparse mat-vec at inference; the physics
anchor is \citet{marzari1997maximally,marzari2012maximally}.
\item[P002 Tight-binding sparse attention.] Here the
expected attention magnitude $\bar A(\tau) = \mathbb{E}\,A_{ij}$ as
a function of token distance $\tau = j - i$ in pretrained
transformers decays exponentially, $\bar A(\tau) \propto e^{-\tau/\lambda}$,
in analogy with hopping integrals in Slater--Koster tight-binding
models. The compression is a distance cutoff at $\tau > 3\lambda$; the physics
anchor is \citet{slater1954simplified,ashcroft1976solid,kohn1996density,prodan2005nearsightedness}.
\item[P003 DMRG-truncated MLPs.] The hypothesis is that per-layer MLP weight
matrices admit an optimal low-rank approximation $W \to W_D$ with $D$
chosen by the eigenvalue spectrum of the input activation density
matrix $C = \mathbb{E}[h h^\top]$, in analogy with White's reduced
density-matrix truncation criterion. The compression is a low-rank factor;
the physics anchor is \citet{white1992density,schollwock2011density}.
\item[P005 Wilson-RG of weights.] The hypothesis is that iterated block-decimation
of pretrained weight matrices, in the spirit of Wilson--Kadanoff
\citep{wilson1971rgI,kadanoff1966scaling}, induces a flow on the
empirical singular-value distribution converging to a universal
fixed-point density $\rho^\star$ that predicts the magnitude-pruning
sparsity at which model quality collapses. The compression is the predicted
sparsity; the physics anchor is \citet{wilson1971rgI,wilson1974epsilon}.
\item[P011 Tensor-train embedding.] The hypothesis is that the per-token bond
dimension $D_v$ of a per-row tensor-train (matrix-product-state)
decomposition of the embedding matrix follows a scaling law
$D_v = D_0\,e^{\gamma H_v}$ in the token surprisal
$H_v = -\log p_v$, so that frequent (low-surprisal) tokens can be
stored at lower rank, in analogy with White/DMRG Schmidt-truncation
under a 1D area law. The compression is a per-token adaptive-rank embedding;
the physics anchor is \citet{oseledets2011tt,white1992density,hastings2007area}.
\end{description}

P001, P002, P003, and P011 are probe-only studies: without training, they
measure attention statistics or decompose frozen weights by SVD/tensor train.
All four reached a pre-registered Phase-3 pilot. In contrast, P005 stopped at
Phase~1, pre-empted by contemporaneous literature
(Section~\ref{sec:p005}) before a pilot received a budget. P001 uses
Pythia-160M \citep{frankle2019lottery}, a small, well-documented GPT-NeoX
model with rotary positional embedding. P002 uses GPT-2-medium (24 layers,
16 heads, learned absolute positional embeddings) and also probes
Pythia-160M to permit the P001--P002 cross-paper consistency check
(Section~\ref{sec:cross_paper}). The evaluation set is the first 10,000
tokens of the \textsc{WikiText-103} test split, with sequence length 512,
stride 256, and perplexity as the primary metric.

\subsection{Pre-registration mechanics}
\label{sec:prereg}

For each Phase-3 paper (P001 and P002), we wrote a YAML pre-registration
containing the hypothesis, model targets, metric, control baselines,
predicted effect size with explicit numerical ranges, and the success and
shelve criteria. We committed these files before collecting any pilot data.
Commit \texttt{10f2330} contains the files
\fpath{pre_registered_hypotheses/20260514_P001_wannier.md} and
\fpath{pre_registered_hypotheses/20260514_P002_tightbinding.md}.
Both pilots ran, and both verdicts were written, between $\sim 21{:}40$ and
$\sim 21{:}55$ Atlantic Standard Time on 2026-05-14, well after the prereg
commit. Thus, the git timestamp on each file provides the immutable mark.

A pre-commit hook (\texttt{prereg\_gate\_hook.py}) was designed to refuse a
\emph{publish event}. Its code defines such an event as a
\texttt{Bash} call matching a paper-build or push pattern
(\texttt{pdflatex}, \texttt{latexmk}, \texttt{git push \dots main}), or a
\texttt{Write}/\texttt{Edit} operation on a manuscript-like path. It refuses
that event if no corresponding pre-registration file has a git timestamp
earlier than the final-experiment timestamp in
\fpath{projects/*/pilot/results.json}. We specify this trigger surface because
earlier versions called it a gate on ``\texttt{paper\_writer\_*}
invocations'', although the code never matched those calls. However, the hook
is not wired, and we find no evidence that it ever was. This phrasing is
deliberate: as measured on 2026-09-22, none of the three settings files loaded
by the project registers it, and the version-controlled file has no record of
that string anywhere in its history. The two user-level settings files are
not under version control, and the hook writes no execution receipt.
Therefore, we cannot support the stronger statement that ``it never ran,
ever,'' and do not make it; we state only that it is not wired now and that no
evidence shows it ever was. We report this as a finding rather than repair it
because, as \S\ref{sec:program} explains, wiring it as written would have
been worse than omitting it. It is the git timestamp that does the work. The
timestamp is a record, not a gate: an honest researcher can still mis-design
the preregistration, but cannot revise a prediction after the outcome without
leaving evidence.

\subsection{The $3\sigma$ gate}

A Phase-3 pilot PASSES only when its headline effect exceeds baseline noise
by at least $3\sigma$ across at least three random seeds. Specifically, P001
required the Wannier-rotation sparsity
$s^\star$ exceed the PCA baseline by $\geq 3\sigma$ in a one-sided
Welch's $t$-test with $n=3$ seeds per arm (equivalent to $t \geq 4.5$
at the $n=3$ effective degrees of freedom). Analogously, P002 required
TB-cutoff perplexity to beat the uniform-window baseline by $\geq 3\sigma$
in a three-way calibration bootstrap.

The shelve rule is deliberately broader than the inverse of the pass rule.
P001 shelves if $s^\star_{\mathrm{Wannier}} < 0.30$
(strictly below the lower edge of the predicted range 0.55--0.75)
\emph{or} if Wannier $\leq$ PCA by any margin \emph{or} if the
cross-layer pattern is reversed. P002 shelves if H2 power-law wins
AIC at the median layer (which directly rebuts the TB framing) \emph{or}
if TB-cutoff loses $\geq 5\%$ perplexity versus full attention. These
asymmetries are intentional. For a cluster in which each paper tests one
specific physical analogy, ``below the predicted floor'' is more informative
than merely ``not significant.''

\subsection{The append-only catalogue}
\label{sec:catalog}

\begin{figure}[t]
\centering
\includegraphics[width=0.95\linewidth]{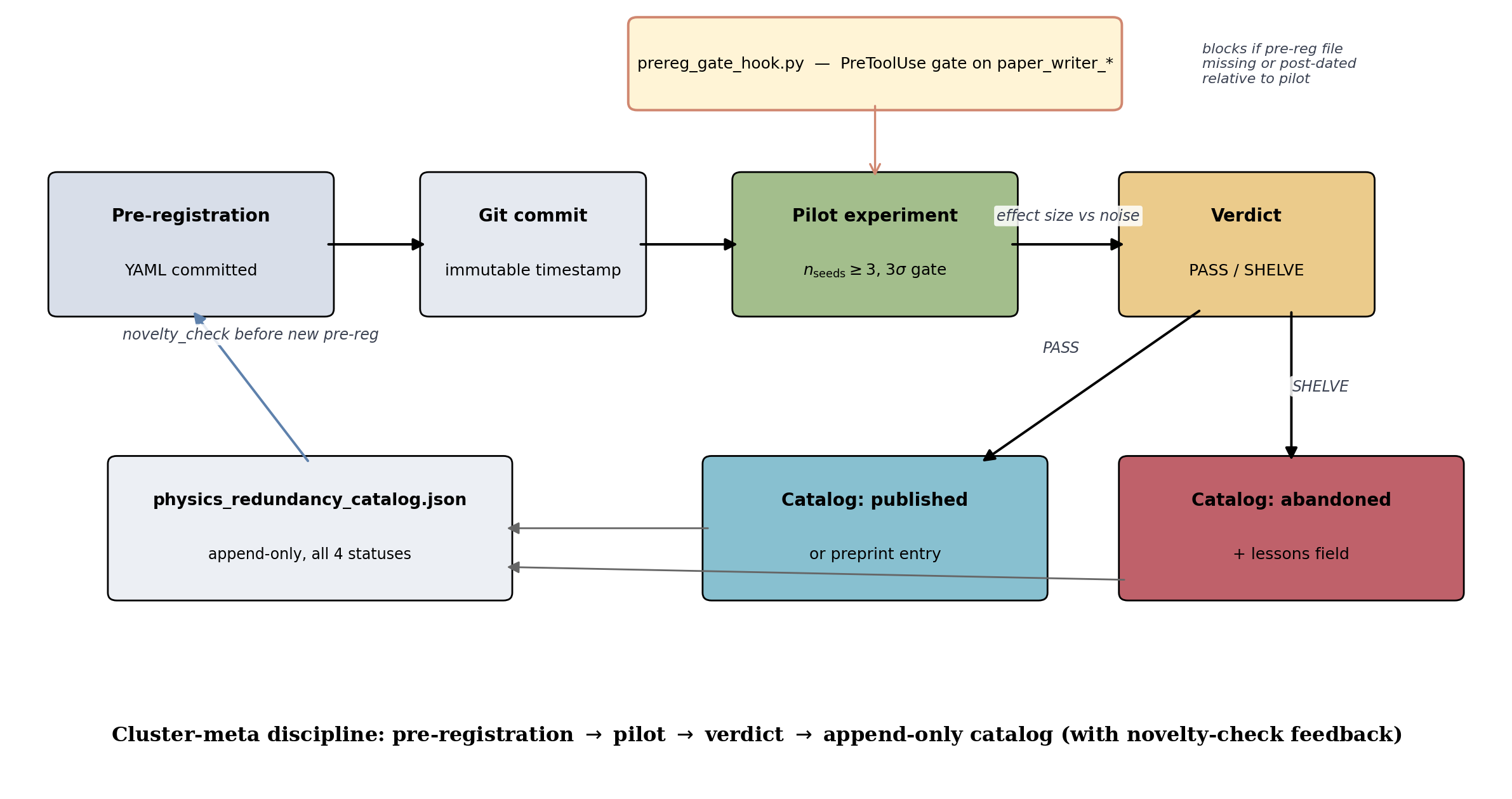}
\caption{Block diagram of the cluster-meta discipline. Committing the
pre-registration YAML to git fixes an immutable timestamp before the
pilot. The pilot uses $n_{\mathrm{seeds}}\geq 3$ and a $3\sigma$
effect-size gate; its PASS / SHELVE verdict creates an entry in
\fpath{physics_redundancy_catalog.json}, either as published/preprint or
as abandoned with a \texttt{lessons} field. Before a pre-registration
is written, a \texttt{novelty\_check} queries the catalogue. It reports
overlap but prevents nothing. The only tool shown that writes
pre-registration files refuses one case---a file with that name already
exists---and never consults the catalogue. Thus, the author reading the
report, not the software, catches a duplicate (physics, redundancy)
attempt. The \fpath{prereg_gate_hook.py} PreToolUse hook (top) was
designed to block build-and-publish events (\texttt{Bash} paper-build or
push commands, \texttt{Write}/\texttt{Edit} on manuscript paths) when a
pre-registration file is missing or post-dated relative to the pilot.}
\label{fig:pipeline}
\end{figure}

Every attempted mapping, whether successful, abandoned, or in progress, is
recorded in \fpath{physics_redundancy_catalog.json}. Each entry contains a
stable identifier, physics anchor, targeted LLM redundancy, prior-art
references, model family, measured gain (if any), compute cost, and a
free-text \texttt{lessons} field. Importantly, abandoned entries remain in
the record. Future agents must consult the catalogue before proposing a
mapping, and a token-overlap novelty check on the \texttt{keywords} field
detects duplicate attempts. The five entries examined here
(\texttt{DMRG\_MLP\_001}, \texttt{WILSON\_RG\_001}, \texttt{TB\_ATT\_001},
\texttt{WANN\_ATT\_001}, \texttt{TT\_EMB\_001}) are summarised in
Table~\ref{tab:catalog}. Figure~\ref{fig:pipeline} shows the full sequence:
commit the pre-registration YAML to git, run the pilot under the $3\sigma$
gate, record PASS/SHELVE in the append-only catalogue, and query the
catalogue through \texttt{novelty\_check} before later pre-registrations.

\begin{table}[t]
\caption{The five catalogue entries reported here. The \texttt{status}
and \texttt{Cost (h)} fields are copied verbatim from
\fpath{physics_redundancy_catalog.json}. After their Phase-3 pilots,
\texttt{DMRG\_MLP\_001} and \texttt{TT\_EMB\_001} both have
\texttt{status:~abandoned}; \texttt{DMRG\_MLP\_001} also has a
documented, catalogue-sanctioned reframe of its surviving cross-paper
finding (Section~\ref{sec:p003}). \texttt{WILSON\_RG\_001} was
abandoned at Phase~1.}
\label{tab:catalog}
\centering
\small
\begin{tabular}{@{}llllr@{}}
\toprule
ID & Physics & Paper & Status & Cost (h) \\
\midrule
\texttt{DMRG\_MLP\_001} & DMRG / reduced $\rho_A$ & P003 & abandoned & 3.4 \\
\texttt{TT\_EMB\_001} & TT / DMRG Schmidt trunc. & P011 & abandoned & 0.6 \\
\texttt{WILSON\_RG\_001} & Wilson RG / block decimation & P005 & abandoned & 0 \\
\texttt{TB\_ATT\_001} & Slater--Koster tight-binding & P002 & abandoned & 0.128 \\
\texttt{WANN\_ATT\_001} & Marzari--Vanderbilt Wannier & P001 & abandoned & 0.299 \\
\bottomrule
\end{tabular}
\end{table}

\subsection{The cluster framing and its falsifiable prediction}
\label{sec:cluster_framing}

The five papers share one intuition from density-functional theory. In
a gapped insulator, the one-particle density matrix decays
exponentially with separation
\citep{kohn1959analytic,kohn1996density,prodan2005nearsightedness,brouder2007exponential,he2001exponential},
making both the basis-localised picture (Wannier) and the
distance-truncated picture (tight binding) cheap. In a metal, both
fail: the density matrix has algebraic tails, and the system supports
collective long-range modes. Most real materials occupy a mixed
regime. The practical answer is to treat the bulk locally but the
correlated subspace expensively
\citep{schollwock2011density,evenbly2009algorithms}.

This common anchor has \emph{two faces}, both associated with the same
exponential density-matrix decay. The distinction matters because the
five papers divide across them. The \emph{distance-locality} face asks
whether attention magnitude decays with token separation. It motivates
P001 (is there a unitary basis in which the head-feature weights are
sparse?) and P002 (does attention decay exponentially with distance?).
Their failures drive the attention-regime inversion in
Section~\ref{sec:synthesis}. The \emph{basis/rank-locality} face asks
whether weight and embedding matrices are cheaply low-rank in a
suitable, possibly tensor-network, basis. It motivates P003 (MLP
weight rank from the activation density matrix), P005 (weight-spectrum
RG fixed point), and P011 (embedding-matrix rank through per-token
tensor-train bond dimension). Kohn nearsightedness makes \emph{both}
faces cheap in an insulator. However, the two faces can fail for different
reasons, and in our data they do (Sections~\ref{sec:synthesis} and
\ref{sec:cross_paper}). Here ``face'' is deliberately loose. The
distance-locality face follows fairly directly from an exponentially
decaying density matrix; the basis/rank-locality face is a
\emph{heuristic extension} to weight and embedding matrices, which
have no literal spatial density matrix. Thus, the split is
organisational and empirical. We do not claim that rank locality
follows rigorously from Kohn nearsightedness. The attention-specific
3a/3b/3c taxonomy below operationalises the distance-locality face; we
return to the rank face in Sections~\ref{sec:p003},
\ref{sec:p011}, and \ref{sec:synthesis}.

The \fpath{projects/_cluster_meta/O_N_feasibility.md} framing document,
committed at \texttt{defbb04} before any pilot, maps this taxonomy to
three regimes for each (layer, head) pair in a pretrained transformer:

\begin{description}[leftmargin=*]
\item[3a. ``Insulator-like'' heads.] Attention magnitude decays
exponentially with token distance, the head-feature weight matrices
are sparse in some unitary basis, and both P001 and P002 succeed.
\emph{Predicted to be the majority}.
\item[3b. ``Plasmon-like'' heads.] These have algebraic decay and
include attention sinks
\citep{xiao2024streamingllm}, induction heads
\citep{olsson2022induction}, and retrieval heads. A distance cutoff
would break the long-range capability they implement. \emph{Predicted
minority, $\sim 5$--$15\%$}.
\item[3c. ``Critical / metallic'' heads.] These are neither local nor
sparsely long-range. This is the most troubling regime.
\emph{Unknown fraction}.
\end{description}

\paragraph{Operationalising the three regimes via P002's AIC fits.}
P002 fits three functional forms to each (layer, head)
attention-vs-distance histogram: H1 pure exponential
$A_0 e^{-\tau/\lambda}$, H2 pure power-law $A_0 \tau^{-\alpha}$, and
H3 stretched exponential $A_0 e^{-(\tau/\lambda)^\beta}$. AIC selects
the winner. H1 is the operational signature of 3a because pure
exponential decay diagnoses the Kohn-nearsighted regime. H2 marks 3b
because pure power-law decay diagnoses the induction-head /
attention-sink / retrieval-head behaviour enumerated as the
algebraic-decay class in cluster-meta §3b. H3 marks 3c because a
stretched exponential with $\beta < 1$ is the canonical diagnostic of
glassy or critical relaxation in disordered condensed-matter systems.

The correspondence H1/H2/H3 $\leftrightarrow$ 3a/3b/3c is operational,
not exhaustive. A power-law-winning head could be a heavy-tailed-RMT
artefact rather than an induction head. Similarly, a
stretched-exp-winning head could mix two exponentials with different
$\lambda$ rather than represent a genuine critical mode. This
ambiguity is part of the price of studying frozen pretrained models.
The map supports the converse claim we need: any (layer, head) in the
3a regime must win H1 on AIC.

\begin{figure}[t]
\centering
\includegraphics[width=\linewidth]{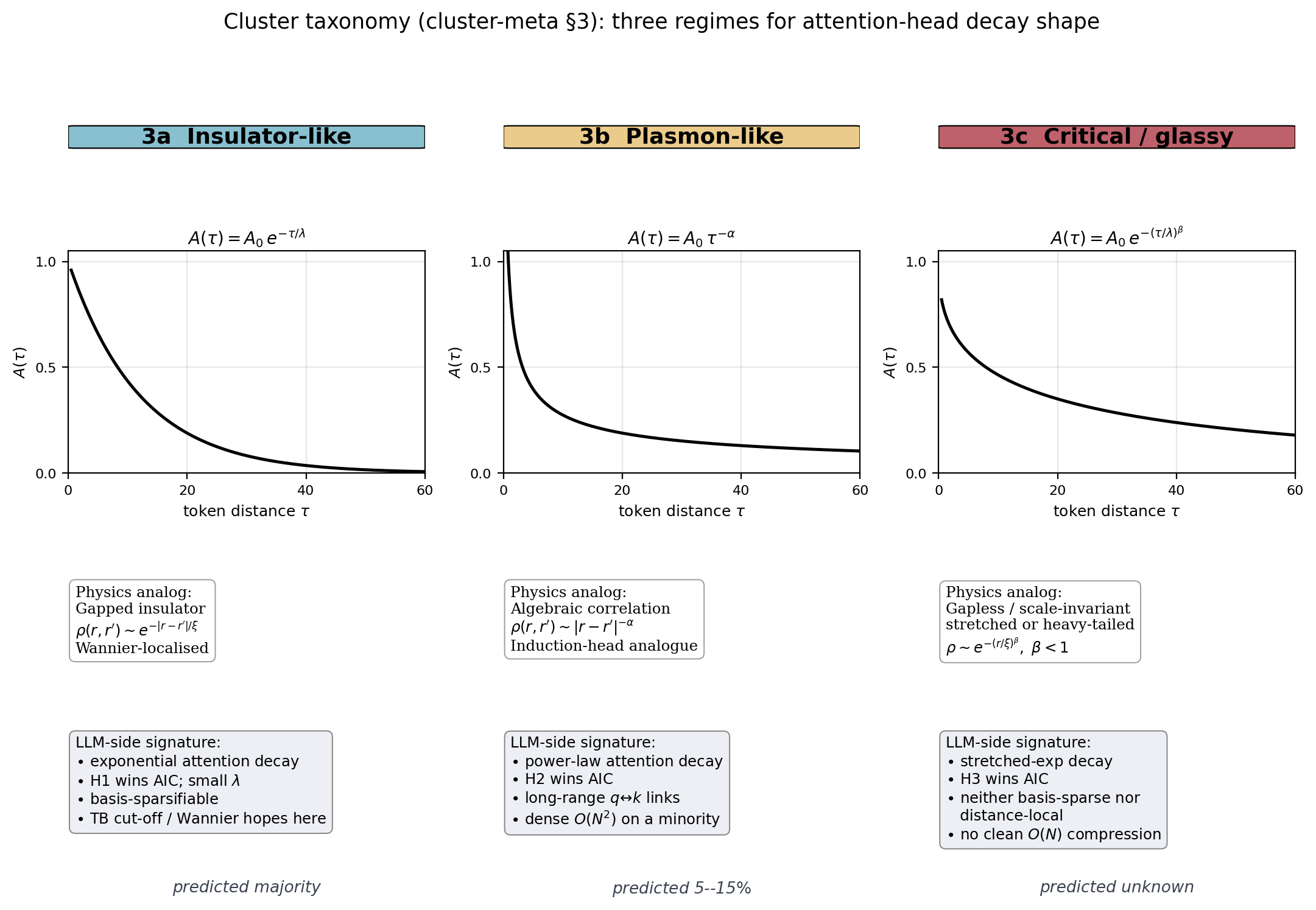}
\caption{Cluster-meta §3 taxonomy of three attention-distance decay
regimes for a (layer, head) pair. Each panel gives the canonical decay
function (top), physical analog (middle), operational LLM signature and
P002 AIC-winner correspondence (bottom), and predicted relative
fraction. The cluster-level prediction makes 3a (insulator-like) the
majority regime; Section~\ref{sec:cluster} shows the measured
inversion.}
\label{fig:taxonomy}
\end{figure}

The cluster-level prediction is therefore that \emph{across pretrained
small transformers, 3a is the plurality; Wannier and tight-binding
compression succeed on most heads, while the few 3b heads can be
handled by an orthogonal mechanism}. Section~\ref{sec:cluster} tests
this prediction against the measured outcomes. Figure~\ref{fig:taxonomy}
compares the three regimes through their canonical decay shapes,
physical analogs, and operational LLM signatures.

\section{Five cluster mappings: four Phase-3 tests and one Phase-1
pre-emption}
\label{sec:tests}

\subsection{P003 — DMRG-truncated MLPs (Phase-3 stage-A pilot)}
\label{sec:p003}

Like the other mappings, P003 began from a physical construction: White's
reduced density-matrix truncation criterion
\citep{white1992density}, which projects the MLP weight $W$ onto the
dominant eigenspace of the input covariance $C = \mathbb{E}[h h^\top]$
(thereby minimising activation-weighted Frobenius reconstruction error).
Up to whitening, this operation is formally identical to the DMRG sweep
step. The Phase-1 kill question asked whether any 2022--2025 LLM-compression
paper had independently derived $W \cdot \mathrm{Chol}(C) \cdot
\mathrm{SVD}\cdot\mathrm{trunc}$ without physics rhetoric. SVD-LLM
\citep{wang2025svdllm} answered yes: it takes the Cholesky factor of
$X X^\top$, computes the SVD of $W S$, truncates, and unscales by $S^{-1}$,
bit for bit the SVD of $W C^{1/2}$ proposed by the DMRG construction.
ESPACE \citep{yu2025espace} reaches the same eigenbasis by projecting
activations, while ASVD \citep{yuan2023asvd} uses the diagonal-only analogue.
Consequently, we dropped the \emph{procedural} claim. P003 instead
pre-registered (commit \texttt{0e37eba}) and piloted at Phase~3 two
falsifiable claims absent from SVD-LLM:
(A) a per-layer scaling law $D_\ell \propto \exp(\alpha S_\ell)$
relating retained rank to the activation-Gram entanglement entropy
$S_\ell$, and (B) an MPO generalisation with the bond dimension $\chi$
set by entropy rather than by hand. The stage-A pilot ran on 2026-06-12
(\texttt{projects/P003\_dmrg\_mlp/pilot/results\_stageA.json}), which
records eight protocol deviations. The material deviation is the model
substitution discussed below. Another matters beyond this paper: the
calibration corpus contained
$\sim\!250$k tokens rather than the pre-registered $1$M, because the
WikiText-103 validation split does not contain $1$M. The same shortfall
independently constrained P011 (\S\ref{sec:p011}). Thus, a pre-registration
fixes not only the protocol but also its assumptions about the dataset. Those
assumptions are themselves falsifiable: here one wrong assumption survived
two independent pre-registrations and external review, and failed only when
confronted with the data.

\paragraph{Registered claim A: shelved.} We fit $\log D_\ell$ against
$S_\ell$ for each layer class on Qwen3-8B.\footnote{The pre-registered
repo-ids were inaccessible at pilot time (Llama-3-8B gated; a
``Qwen3-7B'' does not exist --- the family ships 8B), so Qwen3-8B was
substituted and the deviation logged in the results file. The
quantities we read off the substituted model (singular spectra, the
MPO-reshaping flatness, and the entropy--rank correlation) are generic
properties of trained transformer weight matrices and are not expected
to hinge on the specific $7$--$8$B family, though we did not verify
this across families.} At a $2\%$ per-row reconstruction threshold, the
attention-projection class gives Pearson $r = -0.434$: the \emph{wrong sign}
and below the pre-registered shelve trigger $r < 0.30$. The trigger fired.
The mlp-up class gives $r = +0.502$ and mlp-down gives $r = +0.892$.
However, the mlp-down fit is an own-input-Gram correlation \emph{by
construction}: the retained rank is evaluated against the same covariance
that defines its truncation. It is therefore an exploratory consistency
check, not a compression result, and we do not headline it. An $\alpha$
threshold-dependence downgrade also fired in every class. Thus, registered
claim A is shelved.

\paragraph{Registered claim B (MPO): premise dead at stage-0.} Before any
entropy-driven choice of $\chi$, the MPO ansatz requires a 2-site
matricization of the weight matrix to reveal a \emph{more} compressible
spectrum than the original. It does not. For the exact pre-registered
up\_proj shape (Mistral-7B-v0.3 L16,
$4096 \times 14336 \to 64\times64 / 64\times224$) the balanced 2-site
regrouping needs \emph{more} singular values to capture $98\%$ of the
Frobenius energy than the original matrix (rank-at-$98\%$ 3804 versus
3713). The same is true for Qwen3-8B L18 (3764 versus 3668). The reshaped
spectrum is \emph{flatter}; plain SVD of the original matrix therefore
dominates the MPO factorisation at every parameter budget. Clearly, the
premise fails before the entropy step. P011 fails in the same way
(Section~\ref{sec:p011}): imposing a tensor-network form on a weight object
without genuine 1D locality buys nothing over ordinary SVD. We observe this
three times (P011-GPT2, P011-OPT, P003-B). On the weight side, however, this
is a reproduction rather than a discovery: \citet{zagitov2026rethinking},
posted ten days before the verdict recorded here, reach the same conclusion
from the geometry of approximation error and at $6$--$30$B
(Section~\ref{sec:related}).

\paragraph{Cross-paper check C: passed.} P003 also pre-registered a
consistency check against P002. If both measurements track a
critical/heavy-tailed rather than insulating regime, then layers with more
\emph{stretched-exponential} attention heads (H3, the critical/glassy
diagnostic) should have higher MLP-input activation-Gram von~Neumann entropy
$S_\ell$. On GPT-2-medium, the Pearson correlation between $S_\ell$ and
P002's per-layer H3 fraction is $r = 0.523$ as published, above the pre-registered $r \geq 0.40$ pass threshold --- but $r = 0.0472$ once padded positions are excluded from P002's histograms, far below it. The published value was the cluster's \emph{first passing cross-paper consistency check}; it does not survive the correction, and the positive reading in Section~\ref{sec:positive_reading} loses its support. Even as published the statistical reading was fragile: the check passed under the registered seed-level bootstrap-$t$ analysis ($t = 177.6$), while the more conservative between-layer analytic $z$ was only $2.66\sigma$, below this paper's $3\sigma$ bar, because it assumes likely violated layer independence. Corrected, that $z$ falls to $0.22\sigma$. We therefore report P003-C as failing once the padding defect is corrected, retain the between-layer sensitivity note, and carry it into
Section~\ref{sec:cross_paper} as the \emph{present}-correlation half of a
symmetric pair.

\paragraph{Outcome.} Registered claims A and B shelved; C passed. The
compression window is asymmetric across classes. At the $2\%$ threshold,
the attention $q/k$ projections fall below their break-even rank $D^\star$
in $36/36$ layers of Qwen3-8B, whereas the mlp-down projection does so in
only $8/36$. We therefore set the registered claims in catalogue entry
\texttt{DMRG\_MLP\_001} to \texttt{abandoned}; the surviving P003-C finding
is sanctioned for a documented reframe in a successor paper, whose
as-yet-unrun results do not appear here. One exploratory lesson also entered
our \texttt{PREREG\_CHECKLIST}: the attention class's apparent ``sign
reversal'' arises from pairing $o\_proj$ with a mismatched class Gram.
Against its \emph{own} input Gram, $o\_proj$ gives $r = +0.41$. Thus,
bundling heterogeneous projection types under one Gram is a
pre-registration pitfall, not a physical result.

\subsection{P005 — Wilson RG of weights (Phase-1 pre-emption)}
\label{sec:p005}

The P005 Phase-1 survey applied Wilson's block-spin construction
\citep{wilson1971rgI,wilson1971rgII,kadanoff1966scaling} to a pretrained
weight matrix. The proposal was that iterated $k \times k$ block decimation
would drive the empirical singular-value distribution toward a universal
fixed-point density $\rho^\star$ that predicts magnitude-pruning sparsity. The
RG-meets-deep-learning canon
\citep{beny2013deep,mehta2014exact,lin2017deep,schwab2016comment,roberts2022principles,halverson2021neural,erbin2022nonperturbative,koch2018mutual,li2018neural,levin2007tensor,evenbly2009algorithms}
addresses nearby questions --- learning dynamics, NTK flow, generative
models, and inverse RG --- but does not predict static-weight pruning
density.

Two papers, however, meet P005 directly. \citet{redman2022universality}
formalises iterative magnitude pruning (IMP) as a Wilsonian RG scheme on
weights and reports BERT trajectories ``consistent with RG theory,''
including evidence of flow near fixed points. This pre-empts both the
procedural identification of IMP with RG and the empirical existence of
fixed-point trajectories. Likewise, \citet{lu2024alphapruning} fits
per-layer Hill-$\alpha$ exponents of the empirical spectral density and uses
them to allocate sparsity by layer, pre-empting the allocation half of a
Wilson-RG framing. \citet{martin2018implicit,martin2019traditional}
established the heavy-tailed self-regularisation (HT-SR) framework extended
by AlphaPruning, while \citet{yin2024owl} gives a per-outlier-count
alternative.

In contrast, Mehta--Schwab \citep{mehta2014exact} does not pre-empt the proposal.
That work maps Kadanoff variational RG to stacked RBMs on small Ising
lattices, concerns learning dynamics rather than static pretrained weights,
and does not predict pruning density. Nevertheless, the P005 survey scored
novelty 52/100 (fail). After considering reframes --- cross-model-family
universality of $\rho^\star$ and per-layer RG-flow trajectories not measured
by Redman et al.\ --- we set the catalogue entry to \texttt{abandoned}.

The same pre-emption pattern appears in P003 and P005: a physics-inspired
construction $\to$ independent derivation by a contemporary LLM-compression
paper without the physics rhetoric. The pattern is recorded as
\texttt{LRN-20260514-001}. After the Phase-3 falsifications of P001 and P002,
the pattern appeared a third time at cluster level
(Section~\ref{sec:synthesis}). Thus, \emph{any Phase-1 survey of a ``physics
$\to$ LLM weight-statistics compression'' mapping is incomplete unless it
compares the procedures side by side, as in P003 survey §C.3.1, against
SVD-LLM, AlphaPruning, OWL, and Redman et al.}

\subsection{P002 — Tight-binding sparse attention (Phase-3 falsification)}
\label{sec:p002}

The P002 pre-registration, committed at \texttt{10f2330}, predicted that an
exponential $A_0\,e^{-\tau/\lambda}$ (hypothesis H1) would best fit the
layer-wise attention-versus-distance histogram $\bar A(\tau)$ of GPT-2-medium.
It placed median-layer $R^2$ in the 0.70--0.85 range, made H1 the AIC winner
at that layer, and put the fraction of H1-winning (layer, head) pairs between
0.65 and 0.85. Operationally, the associated TB cutoff at
$\tau \leq 3\lambda$ was to keep perplexity within $0.5$--$2.0\%$ of full
attention and beat a uniform-window baseline by $\geq 3\sigma$ at matched
non-zero count. Optional controls matched BigBird-hybrid
\citep{zaheer2020bigbird} and top-$k$ per-query baselines at the same budget.

The pilot evaluated all $24 \times 16 = 384$ (layer, head) pairs in
GPT-2-medium on 16 distinct documents, duplicated to 32 rows and padded to 512 positions of which a mean 16\% carry real tokens. It fit three competing forms
by nonlinear least squares --- H1 exponential, H2 power law
$A_0\,\tau^{-\alpha}$, and H3 stretched exponential
$A_0\,e^{-(\tau/\lambda)^\beta}$ --- and selected by AIC. The headline
outcomes appear in Table~\ref{tab:p002_headlines}.

\begin{table}[t]
\caption{P002 Phase-3 pilot outcomes versus pre-registered predictions.
All values from \texttt{projects/P002\_tightbinding\_attention/pilot/results.json}, which is padding-inclusive. The AIC-winner row and the three median-layer $R^2$ rows have been recomputed with padded positions excluded, on panel~B (the 4 documents of length $\geq 120$, $\tau \leq 113$); the padded value follows in parentheses. Note the $\tau$ window differs between the two (113 masked versus 128 padded), so the pair is a before/after of the same pipeline rather than a like-for-like refit. Masked values: \texttt{pilot/masked\_median\_layer\_r2.json}.}
\label{tab:p002_headlines}
\centering
\small
\begin{tabular}{@{}lll@{}}
\toprule
Quantity & Pre-registered prediction & Measured \\
\midrule
Median-layer $R^2$, H1 (exp)         & 0.70--0.85       & 0.629 (0.359 padded) \\
Median-layer $R^2$, H2 (power)       & ---              & 0.160 (0.739 padded) \\
Median-layer $R^2$, H3 (stretched)   & ---              & 0.925 (0.801 padded) \\
AIC winner at median layer           & H1               & H3 (12/16); H2 (4/16), masked \\
Fraction(layer,head) H1 wins         & 0.65--0.85       & 0.0052 (2/384) \\
Median $\lambda$ at layer 12         & 24--96 tokens    & 85.6 tokens \\
Full-attention ppl                   & ---              & 23.71 \\
TB at $3\lambda$ ppl                 & ---              & 46.47 ($+96.0\%$) \\
TB ppl $\Delta$ vs full              & $+0.5$ to $+2.0\%$ & $+96.0\%$ \\
Uniform-window matched ppl           & ---              & 12858.5 \\
BigBird matched ppl (mean)           & ---              & 162.7 ($\pm 1.4$) \\
Top-$k$ matched ppl                  & ---              & 23.68 \\
TB beats uniform                     & by $\geq 3\sigma$ & $+8989\sigma$ (yes) \\
TB beats top-$k$                     & within $\pm 0.2$ ppl & loses by $16.0\sigma$ \\
\bottomrule
\end{tabular}
\end{table}

The functional-form result is clear. H1 does not describe GPT-2-medium
attention decay: its median-layer $R^2$ is half those of H2 and H3, and H1
wins AIC in only 2 of 384 (layer, head) pairs, two orders of magnitude below
the predicted floor. Thus, the pre-registered shelve criterion ``H2
power-law wins AIC at median layer'' is triggered.

The operational claim also fails, though the controls matter. At
$3\lambda$, TB beats the uniform-window baseline by an absurd margin
(uniform-window ppl 12858 vs TB 46). This result shows only that a uniform
window is catastrophic for GPT-2-medium attention at the same total budget;
it does \emph{not} support the TB framing. The informative comparison is
against top-$k$ at matched non-zero count. There TB loses by $16\sigma$: the
content-dependent top-$k$ selector retains 23.68 ppl, whereas TB gives 46.47.
Moreover, the absolute TB cost (+96\%) exceeds the pre-registered 5\% shelve
threshold by a factor of 20. The H1 framing fails before the baseline
comparison.

\begin{figure}[t]
\centering
\includegraphics[width=0.95\linewidth]{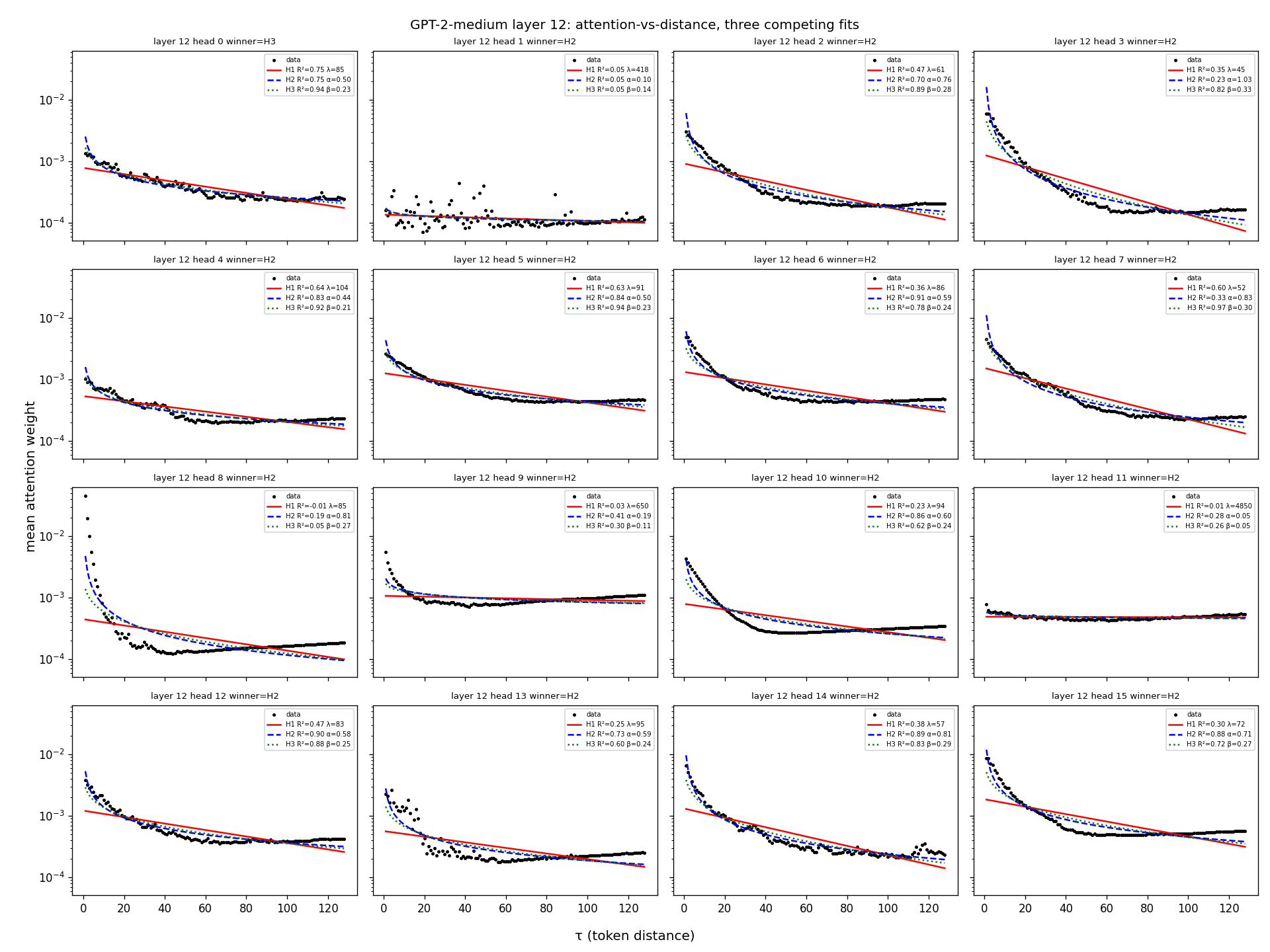}
\caption{P002 attention-versus-distance histograms at the median layer
($L = 12$ of 24) of GPT-2-medium. The 4$\times$4 grid contains all 16 heads,
with the three fits overlaid in each panel: H1 exponential (red), H2 power law
(blue), and H3 stretched exponential (green). Across the 16 heads, H1 undershoots at short
distances and overshoots in the tail; H2 follows the data across the full
$1 \leq \tau \leq 128$ window, and H3 nearly coincides with H2. AIC selects
H2 in 15 of the 16 heads at this layer.
Source: \texttt{projects/P002\_tightbinding\_attention/pilot/figure\_decay\_layer12.png}.}
\label{fig:p002_decay}
\end{figure}

Figure~\ref{fig:p002_decay} overlays the three fits on the median-layer
attention-versus-distance histogram. H1 visibly undershoots at short
distances and overshoots in the tail. H2 and H3 are almost indistinguishable
by eye. Thus, H2 remains the relevant alternative; once the residuals are
this small, AIC prefers H2's single
parameter to H3's two.

\begin{figure}[t]
\centering
\includegraphics[width=0.95\linewidth]{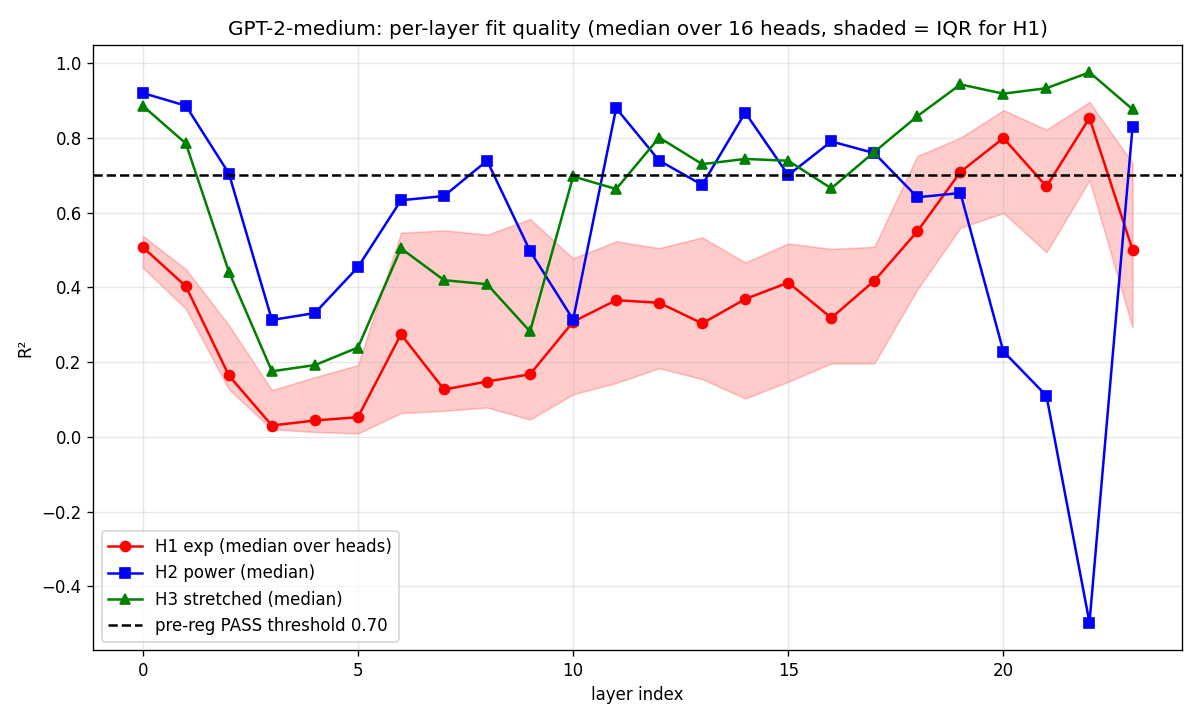}
\caption{P002 per-layer fit quality $R^2$ for H1 (exponential, red), H2
(power law, blue), and H3 (stretched, green) on GPT-2-medium. H2 and H3
dominate H1 at every depth except the final layer, where all three collapse.
Source:
\texttt{projects/P002\_tightbinding\_attention/pilot/figure\_fit\_quality\_per\_layer.png}.}
\label{fig:p002_fitqual}
\end{figure}

\begin{figure}[t]
\centering
\includegraphics[width=0.95\linewidth]{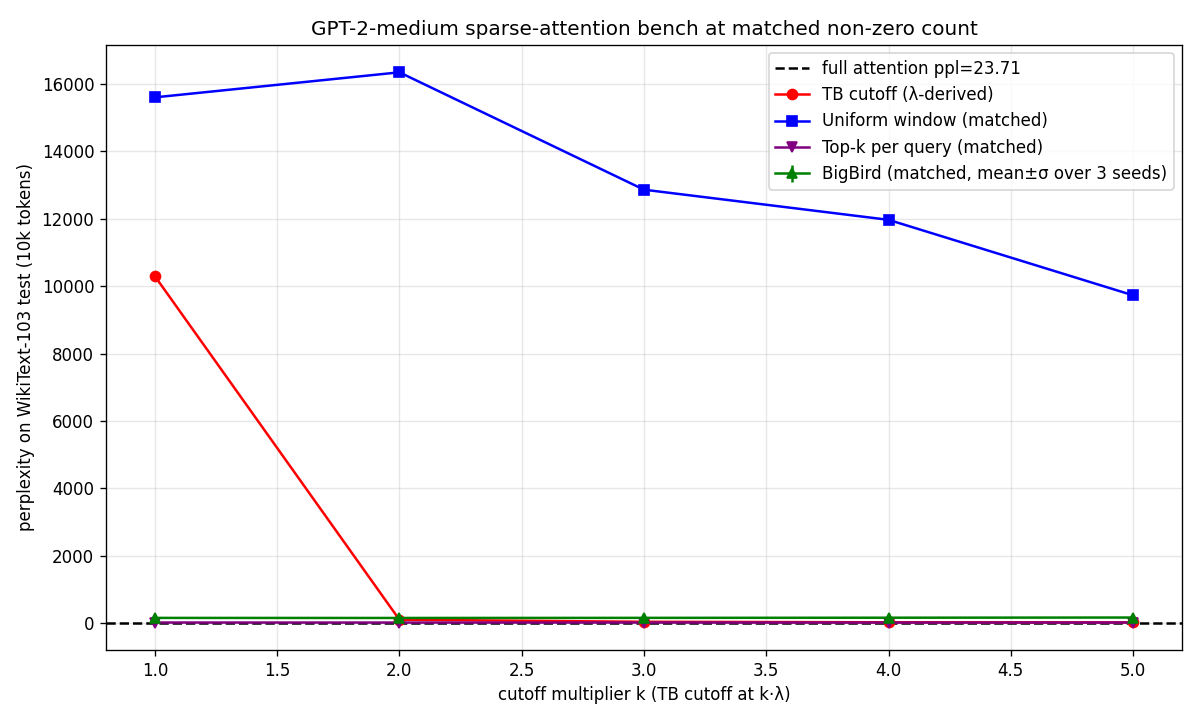}
\caption{P002 perplexity versus TB cutoff multiplier $k$ (cutoff radius
$= k \cdot \lambda$), compared with full attention and matched
uniform-window, BigBird, and top-$k$ controls. TB falls sharply from a
catastrophic value at $k=1$ but does not reach full attention by $k=5$, where
it remains approximately $+15\%$ above the full-attention baseline and well
above the pre-registered $5\%$ shelve threshold. For each $k$, matched
top-$k$ (purple) tracks full attention; the matched uniform window is
catastrophic at all $k$; and BigBird stays near $\sim 160$ ppl. The headline
``TB at $3\lambda$ costs $+96\%$ perplexity'' is the $k=3$ point.
Source: \texttt{projects/P002\_tightbinding\_attention/pilot/figure\_cutoff\_vs\_baselines.png}.}
\label{fig:p002_cutoff}
\end{figure}

The P002 catalogue entry \texttt{TB\_ATT\_001} was set to
\texttt{abandoned} with verbatim lessons: ``(a) GPT-2-medium attention
decay is power-law / stretched, NOT pure exponential. (b) TB analogy
from solid-state insulators incorrectly imports an exponential prior.
(c) Stretched-exp $R^2 = 0.80$ at median layer suggests glassy-regime
physics, not insulator-regime --- future work could re-import via
spin-glass replica framework rather than TB.''

Two prior results make the outcome less surprising. \citet{barbero2025rope}
reject the usual claim that RoPE causes attention to decay with relative
distance on Gemma-7B, confirming that exponential attention decay remains an
open empirical question. \citet{li2025siftattention} fit a power law to the
$\tau$-th quantile of attention scores, although along a different axis from
ours --- per-step quantile evolution rather than a per-distance histogram.
Similarly, \citet{hegazy2025powerformer} impose a smooth heavy-tailed prior on
attention scores in time-series transformers. These results were prior hints
in the direction of our finding.

\subsection{P001 — Wannier-localised attention (Phase-3 falsification)}
\label{sec:p001}

The P001 pre-registration, committed at \texttt{10f2330}, predicted that the
Wannier rotation would reach a maximum sparsity $s^\star$ in the 0.55--0.75
range (central value 0.65) at $\leq 2\%$ perplexity loss on Pythia-160M.
This was well above the predicted 0.35--0.55 range for the PCA baseline and
0.05--0.20 for a random Haar rotation. The construction used the Cayley
parametrisation
$U = (I - A)(I + A)^{-1}$ with skew-Hermitian $A$, optimised by Adam
(\texttt{lr} $= 10^{-3}$, 1000 steps) to minimise the
Marzari--Vanderbilt $\Omega$ functional with a data-induced position
operator $\hat r$ (``Candidate C''), derived from the token-position
autocorrelation of post-attention residual-stream activations.

The pilot evaluated four methods (\texttt{identity}, \texttt{random Haar},
\texttt{PCA}, \texttt{Wannier}) on Pythia-160M
($12 \text{ layers} \times 12 \text{ heads}$), using three random seeds per
method and the standard 10,000-token WikiText-103 test set. The headline
results appear in Table~\ref{tab:p001_headlines}.

\begin{table}[t]
\caption{P001 Phase-3 pilot outcomes versus pre-registered predictions
on Pythia-160M. $s^\star$ is the maximum sparsity at $\leq 2\%$
WikiText-103 perplexity loss; mean $\pm$ std over $n=3$ seeds. Welch
$t$ is one-sided, $n=3$ vs $n=3$. All values from
\texttt{projects/P001\_wannier\_attention/pilot/results.json}.}
\label{tab:p001_headlines}
\centering
\small
\begin{tabular}{@{}lll@{}}
\toprule
Method & Pre-registered $s^\star$ range & Measured $s^\star$ \\
\midrule
identity (sanity floor)   & 0.0          & $0.058 \pm 0.000$ \\
random Haar               & 0.05--0.20   & $0.052 \pm 0.006$ \\
PCA on $W_K W_Q^\top$     & 0.35--0.55   & $0.051 \pm 0.000$ \\
Wannier (Candidate C $\hat r$) & 0.55--0.75 (central 0.65) & $0.054 \pm 0.004$ \\
\midrule
Welch $t$, Wannier $>$ PCA       & passes if $t \geq 4.5$ & $t = 1.16$, $p = 0.187$ (fail) \\
Welch $t$, Wannier $>$ random    & passes if $t \geq 4.5$ & $t = 0.46$, $p = 0.337$ (fail) \\
Mean $\Omega$ mid-layers (4--8)  & smaller than edge & 516.6 \\
Mean $\Omega$ edge-layers (0--3, 9--11) & ---     & 823.5 (mid $<$ edge: true) \\
Baseline ppl (WikiText-103 test) & ---          & 33.6447 \\
\bottomrule
\end{tabular}
\end{table}

Clearly, all four methods collapse to $s^\star \approx 0.05$, including the
\texttt{identity} sanity floor with no rotation. The Wannier result does not
separate from PCA ($p = 0.187$) or random Haar ($p = 0.337$). In every
(layer, head) attention block, removing only the smallest $1$--$2\%$ of weight
magnitudes already produces 2\% perplexity loss, irrespective of basis. The
cross-layer pattern does follow the predicted direction: mid-layers are more
localisable than edge layers by post-optimisation $\Omega$. This was the one
pre-registered prediction to pass. However, it distinguishes only which
heads are \emph{relatively} easier to rotate; it says nothing about
\emph{absolute} sparsity and, in any case, remains below the $3\sigma$
statistical-confidence bar.

\begin{figure}[t]
\centering
\includegraphics[width=0.95\linewidth]{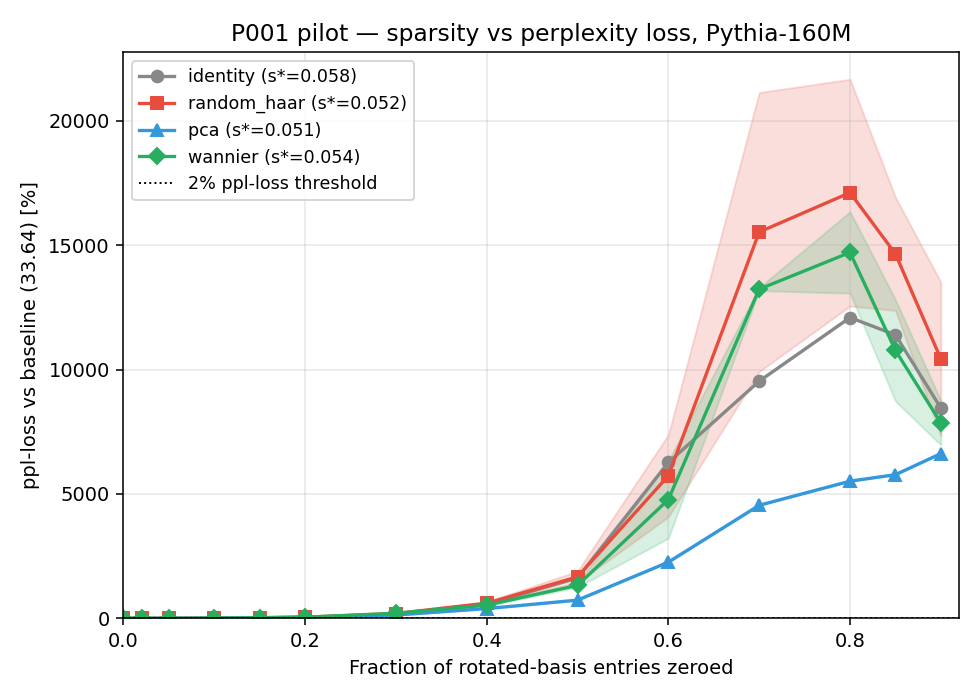}
\caption{P001 perplexity versus sparsity for four rotations on
Pythia-160M, reported as mean $\pm$ std over $n=3$ seeds. All four curves remain
within seed noise at $s \leq 0.3$, and all four reach the $\leq 2\%$
perplexity-loss line at $s^\star \approx 0.05$. The horizontal dashed line
marks the pre-registered $2\%$-of-baseline threshold.
Source: \texttt{projects/P001\_wannier\_attention/pilot/figure\_sparsity\_vs\_ppl.png}.}
\label{fig:p001_sparsity}
\end{figure}

\begin{figure}[t]
\centering
\includegraphics[width=0.95\linewidth]{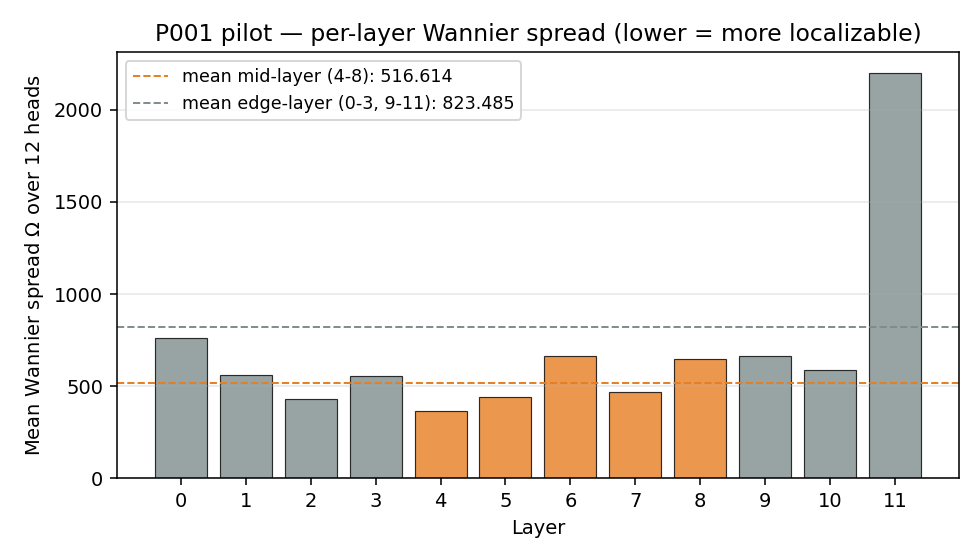}
\caption{P001 per-layer $\Omega_{\mathrm{Wannier}}^{(\ell)}$, averaged over
heads, on Pythia-160M. Mid-layers (4--8) have the lowest $\Omega$; edge layers
(0--3, 9--11) are roughly $1.6\times$ larger. Thus, the pre-registered
direction (mid $<$ edge) is satisfied, but heavy-tailed weight magnitudes,
not basis choice, control the overall $s^\star$ result.
Source: \texttt{projects/P001\_wannier\_attention/pilot/figure\_per\_layer\_sparsity.png}.}
\label{fig:p001_perlayer}
\end{figure}

Figure~\ref{fig:p001_sparsity} shows the perplexity--sparsity trade-off for
the four methods with $n=3$ seed error bands. The four curves overlap within seed noise from
$s = 0$ to $s = 0.3$, then separate idiosyncratically at high sparsity, where
every method lies far above the 2\% perplexity-loss line.
Figure~\ref{fig:p001_perlayer} gives the per-layer post-optimisation
$\Omega_{\mathrm{Wannier}}^{(\ell)}$.

The P001 catalogue entry \texttt{WANN\_ATT\_001} was set to
\texttt{abandoned} with verbatim lessons: ``Wannier rotation with
Candidate C $\hat c$ on Pythia-160M attention weights does not
separate from PCA (or random\_haar) baseline at the
achievable-sparsity-at-2\%-ppl-loss metric. The localisation framing
collapses to structure-preserving rotation on this architecture.
Likely structural cause: a unitary on the head-feature axis is
incompatible with RoPE, which acts on adjacent feature pairs.
Phase~4 NOT committed.'' We discuss the RoPE incompatibility
limitation in Section~\ref{sec:limitations}.

The rotation-compression family
\citep{ashkboos2024slicegpt,ashkboos2024quarot,liu2025spinquant,liu2025larosa}
does \emph{not} target our metric. SliceGPT pursues dimensionality reduction,
whereas SpinQuant suppresses outliers for quantisation; neither reports gains
on the axis ``what sparsity is achievable in the rotated basis at fixed
perplexity?'' Our null result answers a narrower question: \emph{can a
Marzari--Vanderbilt-style $\Omega$-minimising rotation expose pre-existing
latent sparsity?} For Pythia-160M with a Candidate-C position operator, it
cannot.

\subsection{P011 — Tensor-train embedding decomposition (Phase-3 pilot)}
\label{sec:p011}

P011 tests the cluster's rank face on the \emph{embedding} matrix. It is also
the one negative that the frozen embedding's singular spectrum decides
without a single forward pass. The pre-registration (committed
\texttt{97911fb} after two rounds of external peer review) proposed a per-row
tensor-train (matrix-product-state) decomposition with per-token bond
dimension $D_v$ obeying
$\log D_v = \log D_0 + \gamma H_v$ in the token surprisal
$H_v = -\log p_v$. Frequent tokens would then compress to lower rank, giving
$\geq 8\times$ embedding compression at
$\leq 1\%$ perplexity loss, beating a uniform-rank baseline
\citep{hrinchuk2020tensorized} by $\geq 1.5\times$. The physics anchor
is White/DMRG Schmidt truncation \citep{white1992density} under a 1D area law
\citep{hastings2007area}, applied at vocabulary granularity. The tensor-train
format is Oseledets' \citep{oseledets2011tt}; the per-row MPS construction
follows TensorGPT \citep{xu2023tensorgpt}, with entropy controlling the rank
of each row. \emph{Scope of the negative.} We test this \emph{per-row}
construction, reshaping and truncating each embedding row independently.
Methods that decompose the full embedding table as one object
\citep{hrinchuk2020tensorized} use a different construction. They enter the
pre-registration only as a uniform-rank \emph{baseline} to beat, not as a
method we reproduce. Thus, the result bounds per-row MPS with entropy-driven
rank allocation; it is not evidence against tensorized embedding layers in
general. The GPT-2 pilot and OPT-1.3B replication are recorded at
\texttt{projects/P011\_tt\_embedding/pilot/results.json}; headline
outcomes are in Table~\ref{tab:p011_headlines}. That file also records
nine protocol deviations, two of them material. The calibration corpus held
$247{,}289$ GPT-2 tokens rather than the pre-registered $1$M because the
WikiText-103 validation split does not contain $1$M. We therefore used all of
it while preserving the $70/30$ ratio --- a data-availability shortfall, not
a choice. We implemented the $70/30$ split as a contiguous cutoff in token
position rather than a random permutation because the pre-registration also
requires a contiguous $50$k-token perplexity subsample of the fit split,
which a permutation would destroy.

\begin{table}[t]
\caption{P011 Phase-3 pilot outcomes versus pre-registered predictions.
GPT-2 values from
\texttt{projects/P011\_tt\_embedding/pilot/results.json}; OPT-1.3B
from \texttt{results\_opt13b.json}.}
\label{tab:p011_headlines}
\centering
\small
\begin{tabular}{@{}lll@{}}
\toprule
Quantity & Pre-registered prediction & Measured (GPT-2) \\
\midrule
$r(\log D_v^\star,\,H_v)$, eval split & $\geq 0.65$ & $0.016$ ($3\sigma$ CI $[0.003, 0.030]$) \\
Scaling-law slope $\gamma$            & $|\gamma| \gtrsim 0.1$ & $0.0002$ \\
Rows at bond cap $D=16$               & ---        & $94.06\%$ \\
Best compression @ $\leq 1\%$ ppl     & $\geq 8\times$ & $0.644\times$ ($1.55\times$ inflation) \\
\quad advantage vs uniform-rank       & $\geq 1.5\times$ & $1.0\times$ (tie) \\
Robustness $r$ ($H_v$ from train)     & ---        & $-0.004$ \\
Per-row TT params @ $D=16$ vs dense   & ---        & 1193 vs 768 \\
\midrule
                                      &            & Measured (OPT-1.3B) \\
\midrule
Rows at bond cap $D=16$               & ---        & $100\%$ (of $50{,}272$) \\
$r(\log D_v^\star,\,H_v)$             & ---        & undefined (zero variance) \\
Best compression @ $\leq 1\%$ ppl     & ---        & $0.790\times$ ($1.27\times$ inflation) \\
\bottomrule
\end{tabular}
\end{table}

\paragraph{The scaling law has nothing to fit.} For GPT-2, at the $\leq 1\%$ per-row
reconstruction threshold, the optimal per-token bond dimension $D_v^\star$
is almost degenerate: $94.06\%$ of the $50{,}257$ GPT-2 rows require the full
cap $D = 16$, and all others require $D = 15$. A 768-dimensional embedding
row reshaped as $(4,4,4,4,3)$ behaves like a generic full-rank vector. There
is no token-dependent low-rank structure for $H_v$ to predict. Accordingly,
the eval-split Pearson correlation between $\log D_v^\star$ and $H_v$ is
$r = 0.016$
($3\sigma$ Fisher CI $[0.003, 0.030]$; seeds $0.017 / 0.015 / 0.016$),
and the fitted slope is $\gamma = 0.0002$ (seeds
$0.00019 / 0.00018 / 0.00025$). Both pre-registered shelve triggers fire:
$r < 0.30$ and $|\gamma| < 0.1$. Recomputing $H_v$ from $5$M held-out
training tokens rather than the calibration split gives $r = -0.004$.
Therefore, the null does not arise from the surprisal estimate; the
degeneracy lies in $D_v^\star$.

\paragraph{The format cannot compress --- and the spectrum says so.} Two
facts from the frozen embedding matrix settle the question without a forward
pass. First, storage: once the bond dimension reaches $D = 11$ (817
parameters), a per-row TT stores \emph{more} parameters than the dense
768-dimensional row (768), rising to 1193 at $D = 16$. Second, fidelity:
meeting the pre-registered $\leq 1\%$ per-row reconstruction error requires
$D \geq 15$ for almost the entire vocabulary, with $94\%$ of rows at the full
cap $D = 16$. This is a property of the row singular spectra, not perplexity.
Together, the two facts give a best ``compression'' of $0.644\times$, or
$1.55\times$ \emph{inflation}, identical for the entropy schedule and the
uniform baseline. The advantage ratio is $1.0\times$, so even the
``beat-uniform-by-$1.5\times$'' criterion is a tie. Perplexity confirms the
verdict downstream: uniform $D = 15$ already drives proxy perplexity to
$\sim 5.5 \times 10^5$ from a baseline of $28.5$, and only the exact
full-rank allocation is perplexity-neutral. Thus, \emph{row-level linear
algebra alone falsifies the mapping}. This is the cleanest refutation in the
cluster.

\paragraph{Replication sharpens the negative at larger scale.} On OPT-1.3B
($d = 2048$, $V = 50{,}272$), the degeneracy is total:
$100\%$ of the $50{,}272$ rows require the full bond cap at the
$\leq 1\%$ threshold, so $\log D_v^\star$ has \emph{zero variance} and
the Pearson $r$ is formally undefined. The per-token structure required by
the scaling law is absent, not merely weak. This reverses the
pre-registration's premise that a wider embedding
($d = 2048$ versus $768$) would offer \emph{more} exploitable
low-rank headroom. On raw storage the wider row is indeed marginally
closer to break-even: $1.27\times$ inflation at $D = 16$, versus
$1.55\times$ for GPT-2, because the per-row TT overhead is smaller relative
to a 2048-dimensional dense row. But the format still inflates, and the
per-token variation meant to support adaptive rank disappears even more
completely. Widening does not create headroom for the scaling law; it removes
the little present in GPT-2. It is the absence of a chain that breaks the
area-law analogy. The embedding feature axis has no 1D locality: its rows are
unordered coordinate vectors, not a spatial chain. The reshaped-row Schmidt
spectrum is therefore flat, and MPS truncation buys nothing.

\paragraph{An exploratory tied-embedding note.} GPT-2 ties the input
embedding to the output (lm-head) projection. In an exploratory,
\emph{not} pre-registered attribution, we compress one side of an untied
copy. The output arm dominates perplexity sensitivity: input-side
compression at $D = 12$ is perplexity-neutral ($-0.17\%$), whereas
output-side compression at the same or lower rank is catastrophic by many
orders of magnitude. Any future tied-embedding claim must therefore budget
fidelity against the lm-head arm, not the input arm. This is a lead, not a
result.

\paragraph{Outcome.} Both pre-registered gates failed, and catalogue entry
\texttt{TT\_EMB\_001} is \texttt{abandoned}. The transferable lesson is
physical: an area-law / MPS mapping requires genuine 1D locality in the ML
object, and neither an embedding matrix nor a reshaped weight matrix provides
it. Together with P003-B's failed MPO premise, this gives a ``generic-tensor''
no-gain result in three cases (P011-GPT2, P011-OPT, P003-B). Two are on the
embedding side, where the result is ours (Section~\ref{sec:related}). The
lesson therefore becomes Phase-2 question \#1 for every future
MPS/DMRG/area-law mapping: does the ML object contain a 1D chain at all?

\section{Cluster-level synthesis}
\label{sec:cluster}
\label{sec:synthesis}

\subsection{The empirical inversion}

The cluster-meta prediction committed at \texttt{defbb04} made the 3a
``insulator-like'' regime the majority among attention heads in pretrained
small transformers. The four Phase-3 tests jointly reject this picture at
the sampled $\leq 350\text{M-parameter}$ scale. P001 and P002 directly
invert the attention-regime prediction; P003 and P011 independently remove
the corresponding low-rank expectation for weights and embeddings:

\begin{itemize}[leftmargin=*]
\item P002 measures the attention-distance form across
$24 \times 16 = 384$ (layer, head) pairs in GPT-2-medium. H2 power law wins
AIC in $\geq 15$ of 16 heads at the median layer, whereas H1 exponential
wins in only 2 of 384 pairs overall. The attention-distance regime is
power-law rather than exponential: critical-like, not insulator-like.
\item P001 measures basis sparsifiability on the same Pythia-160M attention
grid. No basis --- Wannier, PCA, random Haar, or identity --- reaches more
than $s^\star \approx 0.06$ before $\geq 2\%$ perplexity loss. Heavy-tailed
weight magnitudes dominate; no unitary change of basis localises the relevant
content.
\item The cluster's \emph{rank face} (P003, P005, P011) also fails, now by
direct test rather than literature pre-emption alone. P003's stage-A pilot
shelved its scaling law and killed the MPO premise. P011 showed that no
allocation lets a per-row embedding tensor train compress within its fidelity
budget: it produces $1.55\times$ inflation on GPT-2 and $1.27\times$ on
OPT-1.3B. Meanwhile, existing work already uses the activation-density matrix
(SVD-LLM, ESPACE) and per-layer heavy-tailed allocation (AlphaPruning, OWL).
Across P011-GPT2, P011-OPT, and P003-B, reshaping a weight or embedding matrix
as a tensor network (MPO/TT) produces a spectrum no more compressible than
plain SVD of the original. This ``generic-tensor'' no-gain pattern means that
the reshaped object has no special low-entanglement structure to exploit. It
appears in all three tested reshapings, spanning $0.1$ to $8$B parameters
(Section~\ref{sec:program}); we report three instances, not a universal
theorem. Nor is weight-side priority ours:
\citet{zagitov2026rethinking} establish the same comparison --- tensor
truncation against the matrix-optimal truncation of the same weights
--- theoretically and at $6$--$30$B on dense and MoE architectures, in
a preprint that predates our Phase-3 rank-face verdicts by ten days
(Section~\ref{sec:related}). Our weight-side case is an independent,
pre-registered reproduction at smaller scale. The embedding-side cases remain
ours because their study excludes the embedding matrix throughout. This
result concerns the \emph{rank} face, not the attention regime measured
above. The two faces remain separate.
\end{itemize}

Neither P001 nor P002 succeeds even for a majority subset of heads. P001's
$s^\star = 0.054$ is global over all heads; its cross-layer pattern shows
\emph{relative} variation, but no head reaches the predicted 0.55--0.75
range. In P002, H3 wins AIC for 12/16 median-layer heads once padded positions are excluded (H2 takes the remaining 4), and the 2/384 H1
wins provide a global upper bound across all layers. Thus, 3a is not a
plurality \emph{in any global sense} at this scale. At most, it is a thin
minority.

\begin{figure}[t]
\centering
\includegraphics[width=0.85\linewidth]{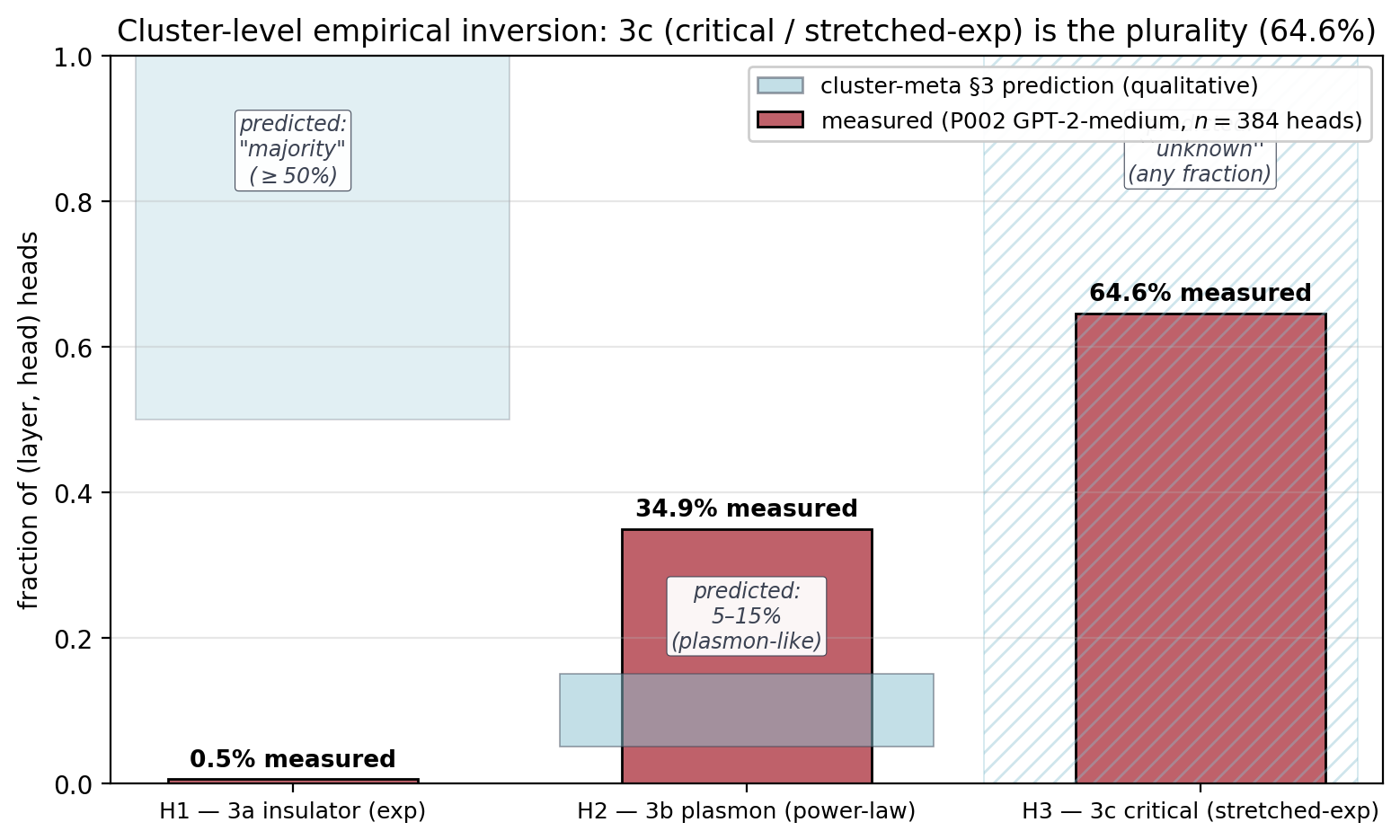}
\caption{Cluster-level empirical inversion. Predicted and measured fractions
of 3a insulator-like, 3b plasmon-like, and 3c critical regimes are
operationalised by P002's competing AIC fits --- H1 exponential, H2 power law,
and H3 stretched exponential --- on GPT-2-medium ($n = 384$ (layer, head)
pairs). The predicted bars render the cluster-meta §3 statement ``3a
majority, 3b 5--15\%, 3c unknown'' as 70/10/20. The data refute the
qualitative inversion, not these exact percentages.}
\label{fig:regime_bars}
\end{figure}

Figure~\ref{fig:regime_bars} condenses the inversion into one plot. The predicted majority-3a distribution becomes, once padded positions are excluded, an empirical plurality of 3c (critical / stretched exponential) at $64.6\%$, with 3b (power-law / plasmon-like) at $34.9\%$ and 3a (insulator-like) reduced to $\approx 0.5\%$. The first draft reported this inversion as a majority-3b distribution ($71.4\%$); that reading was an artefact of the padding and is superseded here, but the direction of the inversion --- away from 3a --- is unchanged and if anything sharper. The 70/10/20 prediction bar is illustrative. The committed
claim was qualitative --- ``3a is majority'' --- and that is the claim the
data falsify.

\subsection{Cross-paper consistency check}
\label{sec:cross_paper}

Both P001 and P002 pre-registered the same cross-paper prediction. On the
shared (head, layer) grid of Pythia-160M, the per-head Wannier spread
$\Omega^{(h,\ell)}$ from P001 should correlate with the per-head TB decay rate
$1/\lambda^{(h,\ell)}$ from P002 at Pearson $r \geq 0.60$, because an
insulator-like head should be local in both basis and distance. If both papers
passed individually but the correlation gave $r < 0.30$, the protocol would
record a cluster-inconsistency event.

\begin{figure}[t]
\centering
\includegraphics[width=0.78\linewidth]{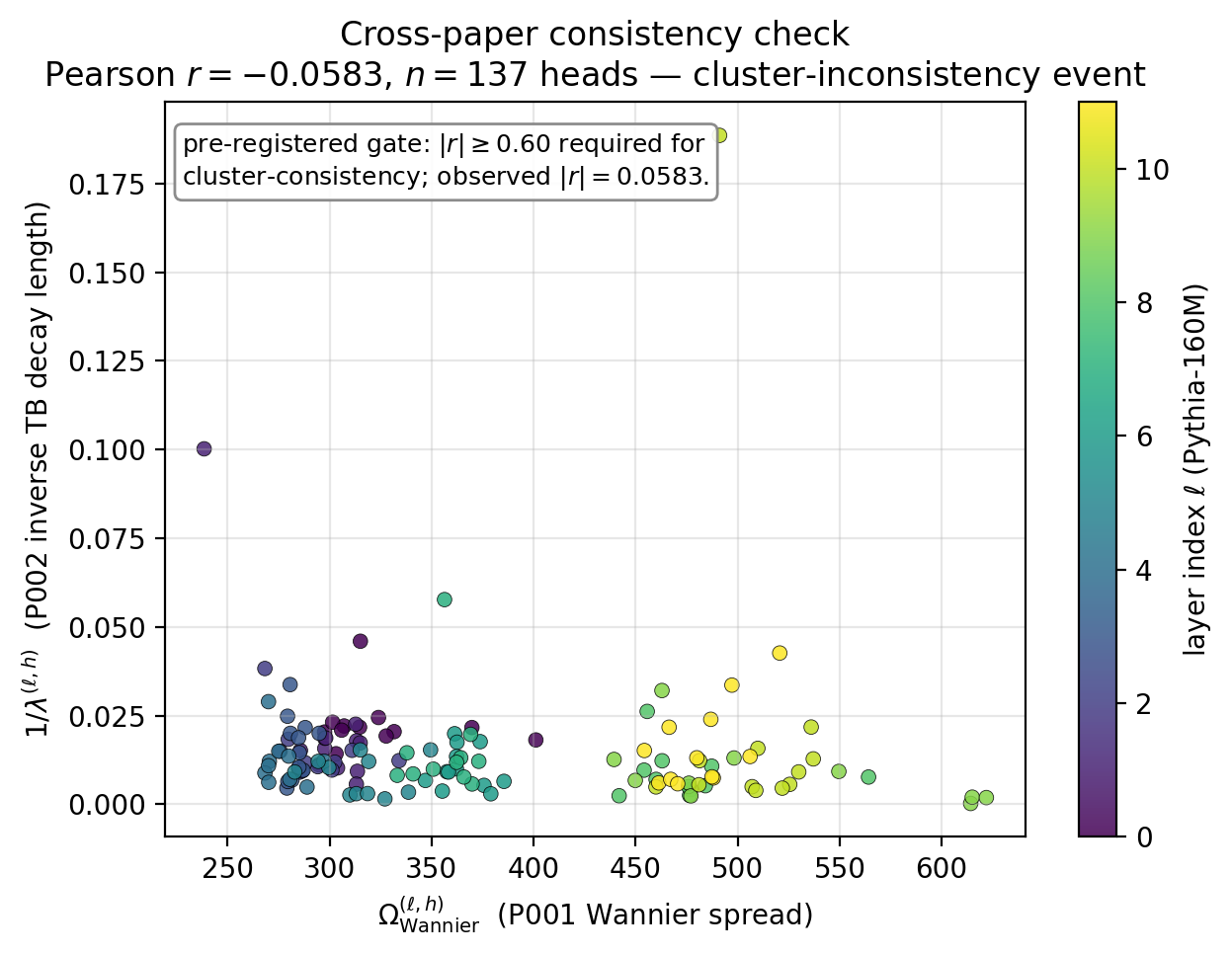}
\caption{Cross-paper consistency on Pythia-160M. P001 Wannier spread
$\Omega^{(h,\ell)}$ (horizontal axis) is plotted against P002 inverse
tight-binding decay rate $1/\lambda^{(h,\ell)}$ (vertical axis) for
$n = 137$ paired (layer, head) cells; colour denotes layer. Pearson $r = -0.0583$, far below the pre-registered $|r| \geq 0.60$ gate. \textbf{This panel has been regenerated with padded positions excluded from both observables}; the padding-inclusive panel the first draft showed ($r = -0.0004$, $n = 121$) is kept as \texttt{figure\_cross\_paper\_scatter\_aspublished.png}. A
cluster-inconsistency event was logged when the P002 pilot concluded.}
\label{fig:cross_paper}
\end{figure}

To make this comparison, P002 probed Pythia-160M in addition to its primary
GPT-2-medium target. Across the paired heads the measured cross-paper Pearson $r$ is \textbf{$r = -0.0583$} ($n = 137$, 95\% CI $[-0.224, +0.111]$); the value published in the first draft was $r = -0.0004$ ($n = 121$), computed before the probe-padding defect was found.\footnote{%
Pythia-160M has $12 \times 12 = 144$ (layer, head) cells in total. \textbf{The published draft described the $n = 121$ figure incorrectly}, as ``explicit removal of the high-$\Omega$ outlier layer 11 (12 heads, on a pre-registered criterion)'' plus eleven NaN drops. No layer was ever removed: \texttt{cross\_paper\_join.py} keeps every cell for which both observables are finite, and all 23 dropped cells fall in layers 3--9 --- layer 11 is present in full, as Figure~\ref{fig:cross_paper} itself shows. The drops are H1 fits whose least-squares slope came out non-negative, so no positive $\lambda$ was defined. Under the corrected (padding-masked) fits only 7 cells drop, in layers 4, 5, 7, 8 and 9, giving $n = 137$. Values: \texttt{projects/P001\_wannier\_attention/pilot/minus0004\_corrected\_r.json} (corrected) and \texttt{projects/P002\_tightbinding\_attention/pilot/cross\_paper\_pearson.json} (as published).}
This value is essentially zero and lies well below the falsification threshold. With $n = 137$ paired heads a two-sided Fisher $z$-transform gives $\geq 80\%$ power to detect $|r| \geq 0.24$ at $\alpha = 0.05$, and the attenuation ceiling set by the measured reliabilities of the two observables ($\rho_{xx} = 0.984$, $\rho_{yy} = 0.995$) is $|r|_{\max} = 0.989$, so the null is a measured null rather than an unresolvable one. The $|r| \geq 0.60$ gate is therefore well within the
detectable range; the $r \approx 0$ result is not a power-starved null. We
logged a cluster-inconsistency event in
\texttt{physics\_redundancy\_catalog.json}.

This inconsistency is informative \emph{independent} of the individual
outcomes. Even if the shelve thresholds were relaxed --- for example, if
P001 at $s^\star = 0.10$ were accepted as a weak positive --- the absent
P001--P002 correlation would still show that basis sparsifiability and
distance localisability do not measure the same mechanism in small
transformers. In an insulator they would, because both arise from exponential
decay of the one-particle density matrix. Their independence here is among
the clearest evidence that the analogy was imported incorrectly.

Figure~\ref{fig:cross_paper} plots the 137 paired $(\Omega^{(h,\ell)}, 1/\lambda^{(h,\ell)})$ values with padding excluded. The cloud is visibly structureless in $1/\lambda$. Masking compresses $\Omega$ from the published $210$--$2707$ to $239$--$622$ and removes the final-layer outliers entirely: $\Omega$ now rises smoothly with depth, from a median of $\approx 280$--$315$ in layers 0--4 to $\approx 480$--$525$ in layers 8--11, while $1/\lambda$ shows no monotone trend against it at any depth. Thus, a few outliers do not produce the near-zero $r$; the two physical observables are genuinely independent in this regime, before and after the padding correction.

\paragraph{A second, passing cross-paper check.} P003's stage-A pilot
(Section~\ref{sec:p003}) tests a \emph{critical}-regime observable rather
than an insulator one. It pre-registered that GPT-2-medium layers with more
stretched-exponential (H3) heads --- P002's critical/glassy diagnostic ---
would have greater MLP-input activation-Gram von~Neumann entropy $S_\ell$.
The measured Pearson correlation was $r = 0.523$ as published, exceeding the pre-registered $r \geq 0.40$ pass threshold; once padded positions are excluded from P002's histograms it is $r = 0.0472$. The symmetry we originally reported therefore does not survive. Both tests are now null: the insulator pair (P001 Wannier spread and P002 decay rate) gives $r = -0.0583$, and the critical pair (P002 H3 fraction and P003 activation entropy) gives $r = 0.0472$. Observables that should track one another under the insulator analogy are indeed independent, but so are those expected to track the critical/heavy-tailed alternative --- so the corrected evidence removes the contrast rather than strengthening the central claim. The published check was fragile even before the correction: $r = 0.523$ passed under the registered seed-level bootstrap-$t$ analysis ($t = 177.6$), while the conservative between-layer analytic $z$ was only $2.66\sigma$, below the paper's $3\sigma$ bar, because it assumes layer independence (Section~\ref{sec:p003}); corrected, that $z$ is $0.22\sigma$. We therefore treat P003-C as a null result, not as corroboration.

\subsection{What the data say positively}
\label{sec:positive_reading}

If the four tests reject the insulator regime, what replaces it? The data
give two consistent positive signals.

\paragraph{Power-law / critical decay in attention distance.}
P002's H2 form is the bare power law $A_0 \tau^{-\alpha}$. H3, the stretched
exponential $A_0 e^{-(\tau/\lambda)^\beta}$ with $\beta < 1$, has a slightly
higher median-layer $R^2$ (0.80 vs 0.74) but loses on AIC because it adds a
parameter. Both imply critical-like rather than gapped exponential decay,
the signature of a gapless / scale-invariant regime. In 1+1D,
conformal-field-theory (CFT) correlators scale as $\tau^{-2\Delta}$, where
$\Delta$ is a scaling dimension; by contrast, $\beta < 1$ in a stretched
exponential diagnoses glassy disorder. The present data do not distinguish
these sources of the observed attention decay.

\paragraph{Heavy-tailed weight-magnitude distribution.}
P001's collapse of all bases to $s^\star \approx 0.05$ diagnoses a
weight-magnitude distribution in which the smallest few percent of weights
carry $\geq 2\%$ of the relevant signal in \emph{every} tested unitary
basis. This is the regime
\citet{martin2018implicit,martin2019traditional} characterise as
heavy-tailed self-regularisation, and it is exactly the regime
AlphaPruning \citep{lu2024alphapruning} and OWL \citep{yin2024owl}
exploit. A unitary change of basis can no more redistribute the $L^2$ norm of
a heavy-tailed distribution than a rotation can sparsify a Cauchy random
variable.

Both signals point away from gapped-insulator physics and toward
\emph{either} critical (CFT-like), \emph{or} glassy (replica-spin-glass),
\emph{or} heavy-tailed-RMT physics. P003-C supplies a third signal in the
same direction and is the first to connect the two faces. In GPT-2-medium,
layers with the greatest fraction of stretched-exponential (H3) heads also had the highest activation entropy as published ($r = 0.523$; $r = 0.0472$ once padded positions are excluded, so this linkage does not survive), nominally linking the attention-side
critical/glassy diagnostic to an independent MLP-side weight-statistics
measurement. The cluster does not distinguish among the three non-insulator
regimes. It does distinguish all three from the insulator analogy.

\paragraph{Honest ranking of the three positive anchors.}
The three candidate anchors are not equally supported. Ranked by the
directness and weight of evidence in the four pilots, they are:
\begin{enumerate}[leftmargin=*]
\item \textbf{Heavy-tailed random-matrix theory} ---
\emph{best-supported.} P001's all-basis collapse to $s^\star \approx 0.05$
directly marks a heavy-tailed weight-magnitude distribution.
\citet{martin2018implicit,martin2019traditional} characterise this regime,
and \citet{lu2024alphapruning,yin2024owl} already use it operationally. Any
successor built on this anchor must therefore clear a higher novelty bar
against AlphaPruning and OWL.
\item \textbf{Critical / power-law decay} --- \emph{also supported, but
connected less directly to physics.} P002's H2 power law wins AIC in
$\sim$71\% of GPT-2-medium (layer, head) pairs, and the attention exponent
$\alpha$ is measurable. A statistical-mechanics literature now gives direct
operational accounts of criticality in attention:
\citet{chen2025criticalattention} prove a $\beta_n \sim \log n$
critical-scaling phase transition for softmax attention;
\citet{poclopez2024dmft} derive a dynamical mean-field theory of
self-attention exhibiting nonequilibrium phase transitions;
\citet{sun2025phase} reformulate the Transformer as an $O(N)$ model
and identify two phase transitions; and \citet{tiberi2024dissecting}
give a statistical-mechanics theory of attention paths. None measures a
power-law exponent for attention decay with token distance, the observable
registered by P002. The CFT $\tau^{-2\Delta}$ mapping therefore remains an
analogy, not a derivation.
\item \textbf{Glassy / replica-spin-glass} --- \emph{least-supported.} Its
only signal is H3's small median-layer $R^2$ advantage over H2 (0.80 vs 0.74),
which H2 reverses under AIC. We retain it because the stretched-exponential
form
$e^{-(\tau/\lambda)^\beta},\,\beta<1$ is the canonical glassy
diagnostic in condensed matter, but the four pilots produce no
direct overlap-distribution or order-parameter signal. A successor choosing
this anchor would need a Parisi-overlap probe, which we did not run.
\end{enumerate}

\subsection{Scale hypothesis}
\label{sec:scale}

A natural objection is scale: the direct tests used only
$\leq 350$M-parameter models. Llama-3-8B and larger models may have different
attention structure; earlier work reports banded patterns
\citep{nawrot2025sparsefrontier}, and the Sparse Frontier benchmark
compares several methods at context lengths up to 128K. We did not run the two
attention-face tests (P001, P002) at $\geq 8$B, whereas the rank-face tests
(P011, P003-B) \emph{did} reach $7$--$8$B and remained negative. Thus, the
open scale question concerns the attention regime specifically
(Section~\ref{sec:limitations}).

The available data permit two readings:

\begin{enumerate}[leftmargin=*]
\item \emph{The inversion is scale-dependent.} Small transformers are
critical / glassy / heavy-tailed, but larger transformers approach an
insulator regime in which Wannier and tight-binding compression succeeds.
\item \emph{The inversion is scale-invariant.} The mechanism producing
power-law attention decay and heavy-tailed weight magnitudes in small
transformers persists at all scales, with at most a finite correction; the
insulator analogy is generally wrong for transformer attention.
\end{enumerate}

Distinguishing (1) from (2) requires exact replications of P001 and P002 at
$\geq 8$B, with RoPE, ALiBi, and learned-absolute controls to exclude
positional-encoding artefacts. Two later pre-registered studies ran
\emph{part} of that experiment. Which part matters, because it determines
what those studies can settle.

\emph{What was run:} P013 extended the P002-style attention-decay measure
along a within-family Pythia ladder from $14$M to $12$B; P014 tested transfer
of the decay-form assignment across Pythia, OPT, and GPT-2 at matched scale.
\emph{What was not:} P001, the Wannier-localisation arm, has no
$\geq 8$B replication, and no study contains an ALiBi arm. The
positional-encoding control is therefore limited to rotary and
learned-absolute forms, and the stated discriminating experiment remains
unrun. We include P013 together with the other study because they change the permissible claim and
because the programme requires every concluded negative to enter the record
(Section~\ref{sec:catalog}).

\paragraph{P013: the ladder reaches 12B, and the registered order
parameter was the wrong one.} P013 extended the attention-decay measurement
along a within-family ladder to $12$B. For its registered order parameter
$f_{H1}$ --- the insulator-like fraction, which reading~(1) predicts should
grow with $N$ --- the result is plainly negative: $f_{H1}$ never exceeds
$0.08$ at any rung. On this observable, no accessible scale shows an
insulator regime. However, this does not kill reading~(1). P013 itself found
this to be the weakest of the three available observables, and its
confirmatory arm returned \texttt{AMBIGUOUS}. A negative from a poor order
parameter constrains reading~(1), but does not close it.

This is the weaker half of P013. The stronger half concerns the registration:
$f_{H1}$ has the \emph{smallest} span
($0.071$) and the \emph{weakest} monotonicity ($\rho = -0.56$) of the
three simplex components, and it was locked before any data existed.
Its scaling is not monotone at all --- $0.0772$ at $14$M, $0.0063$ at
$1.4$B, rising again to $0.0289$ at $12$B --- and a quadratic beats the
registered monotone log-linear form by $\Delta\mathrm{AIC} \approx 9$,
with $99.9\%$ of bootstrap draws convex and a vertex at
$N \approx 1.35 \times 10^{9}$ ($95\%$ CI $6\times10^{8}$ to
$2.4\times10^{9}$). The non-monotonicity is \emph{exploratory}: the
registration tested a monotone trend, so the better curve found later is a
hypothesis, not a result. P013's confirmatory arm returned
\texttt{AMBIGUOUS}.

\paragraph{P014: the scale story was a family story.} P014 asked whether the
decay-form assignment transfers across model families at matched scale. It
does not. A within-family scaling effect replicates under corpus perturbation
($z = -0.38$ out-of-sample), yet \emph{inverts in sign} across families at
the same parameter count. The registered claim was falsified, and the
catalogue records the study as abandoned with no measured gain.

This is the relevant result here, and it is methodological rather than about
attention: \textbf{the preceding question is mis-specified.} Readings (1)
and (2) both make scale the axis on which the insulator analogy survives or
fails. Either can be answered only with family fixed, a condition established
by neither us nor the literature under discussion. Replication inside one
pipeline is not independence. P014 makes the problem especially clear in its
own amendment: GPT-2 was justified as an independent family partly
\emph{because} it uses learned-absolute rather than rotary position
embeddings, yet the other ``independent'' family, OPT, also uses
learned-absolute embeddings. The stated reason for independence is itself a
shared property.

\paragraph{What we do and do not claim from P014.} P014's most interesting
output --- that position-encoding family, rather than parameter count,
organises attention decay --- is \textbf{unregistered and exploratory}. It
appeared only after the registered rule returned its verdict, and the
programme's own prescription
(Section~\ref{sec:discipline}) is that a differentiating argument
retreated to after a registered claim fails must be held to the same
standard as the original. We therefore claim neither a finding nor priority:
\citet{dangal2026attention} independently reached the same reading from a
different direction, with an explicit mechanism, in a preprint submitted
approximately one month before P014 concluded. Our registered machinery adds
something narrower --- functional-form fitting with selection among candidate
decay laws, discrete decay-shape classes, a systematic within-family scale
ladder, and pre-registration --- none of which appears in that work.

Thus, the earlier commitment of this section stands, but for an unexpected
reason: \emph{the cluster-meta majority-insulator prediction was specific to
small-LM scale, was the pre-registered prediction tested, and is the one the
data falsify.} The $\geq 8$B extension is no longer a question we simply
decline to answer. Its registered order parameter answered negatively, and
the test exposed a family confound that makes the original two-reading frame
the wrong question. Any future proposal of insulator-like physics for LLM
attention must address both results before claiming novelty.

\section{What this implies for the program}
\label{sec:program}

\subsection{P009 stays gated}

The cluster-meta frame was operationally decisive for a fifth paper, P009
(DFT-functional attention). P009 would replace the
$\mathcal{O}(N^2)$ pairwise attention softmax by an $\mathcal{O}(N)$ local
functional $F[\rho_\text{local}]$ of activation density. Because its training
requires cloud-burst compute, submission was explicitly gated on the joint
P001--P002 outcome: ``do not submit until P001 and P002 jointly show that
$\geq 80\%$ of (layer, head) pairs satisfy 3a (exponential decay AND
basis-sparse).''

Neither condition holds. Consequently, the pre-registered gate remains
closed, and P009 is deferred. P009's deferral inherits the scale caveat of
Section~\ref{sec:scale}: the gate failed at $\leq 350$M-parameter scale, not
universally. To reopen it, a successor must perform exactly the
$\geq 8$B-scale P001--P002 replication described in
Section~\ref{sec:scale}.

\subsection{Recommended physics-anchor pivots}

The positive signals in Section~\ref{sec:positive_reading} suggest three
physical anchors for a successor in this cluster:

\begin{description}[leftmargin=*]
\item[Replica spin-glass theory.] If the regime is glassy, as suggested by
H3's $R^2 = 0.80$ and P001's all-basis collapse, the natural tool is Parisi's
replica-symmetry-breaking ansatz. The corresponding LLM observable is the
per-(layer, head) overlap distribution across calibration samples.
\item[Conformal field theory / critical decay.] If the regime is critical,
as suggested by H2's clean power-law fit, the natural tools are CFT scaling
exponents and operator-product expansion structure. The corresponding LLM
observable is the attention-decay exponent $\alpha^{(h,\ell)}$ and its
dependence on relative depth, context length, and tokeniser geometry.
\item[Heavy-tailed random matrix theory.] If the regime is heavy-tailed, as
suggested by Martin--Mahoney HT-SR and P001, an established toolkit already
exists in the LLM-compression literature
\citep{martin2018implicit,lu2024alphapruning,yin2024owl}. A successor must
therefore clear a higher novelty bar against AlphaPruning and OWL.
\end{description}

We do not select among these anchors here. We recommend only that a future
compression-cluster proposal either (a) reach $\geq 8$B at Phase 3 for an
\emph{attention}-regime claim --- the present inversion is small-LM-only,
and these five mappings cannot determine whether, or in what form, it
persists at $\geq 8$B --- or (b) begin from one of the regimes above rather
than insulator / area-law physics. The \emph{rank} face already permits a
stronger statement. The generic-tensor no-gain result
(Sections~\ref{sec:p003} and \ref{sec:p011}) holds up to $7$--$8$B (P003-B
on Mistral-7B-v0.3 and Qwen3-8B; P011 on OPT-1.3B). Thus, ``test at larger
scale'' does \emph{not} rescue tensor-network reshaping of a weight or
embedding matrix. This partially answers the scale objection for the rank
face. On the weight side, the answer is overdetermined:
\citet{zagitov2026rethinking} reach it at $6$--$30$B. It is ours alone
only for the embedding matrix, for which our ceiling is OPT-1.3B.

\subsection{Program-level changes}

The cluster outcome calls for three concrete programme changes:

\begin{enumerate}[leftmargin=*]
\item \emph{Compression-cluster scale ratchet.} Before claiming
insulator-regime physics, a successor in the compression cluster must run its
falsification test at $\geq 8$B parameters. The extra compute removes the
failure mode demonstrated here: falsification only at small-LM pilot scale.
\item \emph{Procedural-equivalence side-by-side math is mandatory.}
Twice in this cluster, a 2022--2025 LLM-compression paper independently
derived the proposed physics-anchored procedure without its rhetoric:
SVD-LLM for DMRG and Redman+AlphaPruning for Wilson RG. Every Phase-1 survey
must therefore compare the proposed construction explicitly and side by side
with SVD-LLM, AlphaPruning, OWL, and Redman et al.\ before claiming novelty.
P003 survey §C.3.1 provides the template.
\item \emph{MPS/DMRG/area-law mappings need a 1D locality first.} The
central transferable lesson of the rank-face tests is that tensor-network
(MPO/TT) reshaping buys nothing over plain SVD when the ML object has no
genuine 1D locality to truncate. We observed this three times (P011-GPT2,
P011-OPT, P003-B), and \citet{zagitov2026rethinking} found it independently
for weight matrices. Any future MPS/DMRG/area-law proposal must therefore
answer Phase-2 question \#1 --- \emph{is there a 1D chain in the ML object at
all?} --- before pre-registering a compression claim.
\item \emph{Negative-result publishing is a habit, not an afterthought.}
This paper is the proof of concept: five shelved subpapers become one
publishable cluster-level methodology paper. Future cluster outcomes,
positive or negative, should follow the same protocol.
\end{enumerate}

\section{Related work}
\label{sec:related}

\paragraph{Physics-AI mappings.} Work connecting RG and deep learning
\citep{mehta2014exact,lin2017deep,schwab2016comment,beny2013deep,roberts2022principles,halverson2021neural,erbin2022nonperturbative,koch2018mutual,li2018neural,levin2007tensor,evenbly2009algorithms}
asks how physical structure appears in deep networks and how RG intuition
accounts for scaling or training dynamics. None performs cluster-level,
pre-registered tests of compression mappings. Our contribution is
orthogonal: it is a discipline for testing any candidate mapping, not a new
mapping.

\paragraph{Rotation-compression family.} SliceGPT
\citep{ashkboos2024slicegpt} combines an orthogonal transform $+$
dimensionality reduction to compress LLaMA. QuaRot
\citep{ashkboos2024quarot} uses a Hadamard rotation for 4-bit quantisation;
SpinQuant \citep{liu2025spinquant} learns the rotation; and La RoSA
\citep{liu2025larosa} applies layerwise rotated sparse activation. All four
methods seek either dimensionality reduction or outlier suppression for
quantisation. None asks the P001 question: ``can a Wannier-style
$\Omega$-minimising rotation reveal pre-existing latent sparsity at fixed
perplexity?''

\paragraph{Per-layer allocation family.} AlphaPruning
\citep{lu2024alphapruning} fits Hill-$\alpha$ exponents by layer and allocates
sparsity accordingly; OWL \citep{yin2024owl} allocates by outlier ratio; and
Liao et al.~\citep{liao2024layercollapse} show that high-sparsity,
unstructured pruning can shorten effective computational depth (``layer
collapse''). These methods now occupy the niche targeted by P005. As
Section~\ref{sec:program} recommends, any successor must test explicitly
against AlphaPruning and OWL.

\paragraph{Sparse-attention family.} Sparse Transformer
\citep{child2019generating}, Longformer \citep{beltagy2020longformer},
BigBird \citep{zaheer2020bigbird}, Reformer
\citep{kitaev2020reformer}, Linformer \citep{wang2020linformer},
Performer \citep{choromanski2021rethinking}, Routing Transformer
\citep{roy2021routing}, FlashAttention \citep{dao2022flashattention,dao2024flashattention2},
state-space models \citep{gu2022s4,gu2023mamba,poli2023hyena},
StreamingLLM \citep{xiao2024streamingllm}, and the Sparse Frontier
benchmark \citep{nawrot2025sparsefrontier} provide the operational context
for P002. BigBird and Longformer are closest to our matched-budget control.
On GPT-2-medium, however, the matched uniform window underperforms matched
top-$k$ by orders of magnitude. This observation stands independently of the
TB framing.

\paragraph{Current sparse-attention SOTA (2025--2026).} Practical
long-context sparse attention has moved well beyond the 2020--2021 controls
above. Native Sparse Attention (NSA)
\citep{yuan2025nsa} achieves $100\%$ needle-in-haystack accuracy at
$64$K with up to $11.6\times$ decoding speedup and is deployed in
DeepSeek-V3.2. MoBA \citep{lu2025moba} applies mixture-of-experts gating to
attention blocks and runs in production in Kimi. Kimi Linear
\citep{zhang2025kimilinear} reports that a linear-attention hybrid
beats full MLA at 48B with $6\times$ decoding throughput at $1$M
context. \citet{shen2025ssa} (the
arXiv ``Sparse Sparse Attention'') proves that approximation error
scales linearly with attention mass dropped, a theoretical sibling
of P001's $s^\star$. \emph{This paper does not benchmark against any of these
methods}: it proposes a discipline rather than a sparse-attention
construction. We cite them to locate that contribution. The discipline is
meant to govern any future sparse-attention or compression proposal,
including this frontier.

\paragraph{The two ``SSA''s (acronym disambiguation).}
Two recent works share the acronym ``SSA'' and can be confused.
\citet{shen2025ssa} (arXiv:2511.20102, ``Sparse Sparse Attention'') give a
training framework that aligns sparse- and full-attention outputs in feature
space and prove that approximation error grows \emph{linearly} with the
dropped attention mass. This clean theoretical quantity is a sibling of
P001's $s^\star$, the maximum sparsity at fixed perplexity loss, and any P002
successor must use it as a baseline. Separately, Subquadratic Inc.\
(\texttt{subq.ai}) released a May 2026 blog post attributing the
same acronym to ``Subquadratic Sparse Attention.'' As of writing, this second
work is a marketing post without a paper, released code, model card, or
disclosed authors. Its construction cannot be evaluated, and the trade press
has explicitly called for independent replication
\citep{venturebeat2026subquadratic}. We do not treat it as analytical prior
art and do not repeat its headline numbers. We mention it only to separate
the acronym from \citet{shen2025ssa} and to mark a future probe if weights
become available.

\paragraph{Attention decay priors.} Three contemporaneous papers impose
specific decay priors. Gradformer
\citep{liu2024gradformer} imposes exponential decay on graph
transformers; Radial Attention \citep{li2025radial} imposes radial
masks with energy-decay in video diffusion; Powerformer
\citep{hegazy2025powerformer} imposes heavy-tailed (power-law-like)
priors on time-series attention; SiftAttention
\citep{li2025siftattention} fits power-laws to quantile-evolution
during generation; Selective Attention
\citep{neurips2024selective} connects sparsification to
$\kappa \approx n^{-\alpha\gamma}$. Shaham et al.\
\citep{shaham2025molecular} impose power-law biases on molecular
transformers. None tests the decay shape of pretrained-LLM attention directly
through a per-distance histogram, three competing fits, and AIC selection.
This contribution by P002 accompanies P002's negative TB result.

\paragraph{Tensor-network compression of LLMs.} On the rank-truncation axis,
MPO compression
\citep{liu2021enabling,gao2020compressing,javanmard2026compressing}
and SVD-LLM \citep{wang2025svdllm} represent the state of the art.
Stoudenmire--Schwab
\citep{stoudenmire2016supervised} is the canonical precedent for
using DMRG sweeps in an ML pipeline. As Section~\ref{sec:p003} explains,
SVD-LLM pre-empts P003's original procedural claim. More directly,
\citet{zagitov2026rethinking} compare tensor truncation with the
matrix-optimal SVD truncation of the same weights and reach our rank-face
result with stronger support. Their Proposition~1 shows that reshaping
$\varphi$ preserves the Frobenius norm but not the operator norm:
$\|W\|_2 / \|\varphi(W)\|_\sigma \in [1, \sqrt{\min(m,n)}]$, and the
upper bound is generically attained for delocalised singular vectors, so
no tensor analogue of the Eckart--Young--Mirsky theorem exists. Their
Corollary~1 makes this spectral-to-Frobenius gap binding at practical ranks
for the heavy-tailed spectra of transformer weights. Their experiments cover
GPT-J 6B, LLaMA-2 7B, Qwen3-30B-A3B, and GPT-OSS-20B, including dense and MoE
models. The preprint is dated 2026-06-02, ten days before our Phase-3
rank-face verdicts; we found it only in September 2026 through a footnote in
a later survey. We therefore claim no weight-side priority and report P003-B
as an independent reproduction at smaller scale. Our result adds two points.
First, their storage accounting always excludes embedding parameters and
never treats the embedding matrix as a compression target. P011's
embedding-side negative, including the $100\%$-saturated bond dimension that
makes its registered scaling law undefined, is therefore outside their
scope. Second, the two arguments are independent: theirs concerns approximation
geometry; ours concerns the entanglement structure assumed by the physics
mapping. A successor must overcome the latter claim.

\paragraph{Surveys and the compression-realization gap.}
\citet{tarasov2026tensor} survey tensor methods over a seven-stage LLM
lifecycle and define $\rho_{\rm gap}$, the ``compression-realization gap''
between theoretical memory reduction and measured system-level speed. Their
frame is adjacent to ours, but our statement that tensorisation ``buys
nothing over plain SVD'' is not a restatement of that gap.
$\rho_{\rm gap} = C_B/S_T$ places compression in the numerator and requires
matched dense/TN quality for comparison; it asks whether savings that
\emph{exist} become wall-clock gains. Our rank-face result precedes that
question. At the required reconstruction fidelity, there are no savings to
realise: the best ``compression'' is $0.644\times$, i.e.\ $1.55\times$
inflation. The mapping therefore never enters the $\rho_{\rm gap}$ frame.
The two works also differ in kind. That survey reports no new measurements and
builds its gap table from the limitations sections of surveyed papers; the
present paper is a pre-registered, $3\sigma$-gated, cluster-level
falsification. A footnote in that survey also led us to
\citet{zagitov2026rethinking}.

\paragraph{Random-matrix analysis of transformer weights.}
\citet{staats2025small} find task-relevant information in small singular
values, cautioning against naive SVD truncation. \citet{thamm2025pruning}
prune ViTs by Marchenko--Pastur deviation. Together, these works provide the
random-matrix context for the heavy-tailed regime diagnosed by P001.

\section{Methodology contribution: a discipline}
\label{sec:discipline}

Section~\ref{sec:method} presents the method; here we place it in
context and add one engineering note. Pre-registration has a long
history in clinical trials and psychological science
\citep{nosek2018preregistration,chambers2014registered}. AsPredicted,
the OSF Registries, and Registered Reports codify the practice. It has
also entered machine learning in two forms. The NeurIPS 2020
pre-registration workshop and its successor
\citep{bertinetto2021preregistration} adopted the registered-report
model, reviewing an experimental protocol before results exist.
\citet{hofman2023preregistration} instead propose a lightweight
template for predictive modelling that fixes the research question,
dependent and independent variables, training and test data,
transformations, metrics, and baselines before the test set is touched.

Our contribution is neither pre-registration nor its transfer to
machine learning. It is an adaptation to physics-AI mapping research
with five operational additions: (i) a $3\sigma$ effect-size gate as a
hard PASS/SHELVE rule rather than a $p$-value threshold; (ii) an
append-only catalogue against which future agents check both our past
attempts and the prior art recorded by earlier agents on matching
entries---a deduplication gate over recorded knowledge, not a
literature search; (iii) cluster-meta framing that turns $N$
individual papers into one $N$-test cluster experiment; (iv)
cross-paper consistency checks between physical observables expected
to correlate under the cluster prediction; and (v) daemonic execution
under pre-registration discipline, including a measured account of why
its mechanical enforcement was designed but \emph{not} deployed for
the LLM/MCP-tooling era of autonomous research agents.

Relative to \citet{hofman2023preregistration}, the load-bearing
additions are (i), (ii), and (v). Their template is a socially
sustained document norm. We attempted to check the same norm
mechanically at tool-call time and could not. The gate is registered
in none of the three settings files loaded by this project and is not
wired now (\S\ref{sec:prereg}). Wiring it as written would be worse
than omitting it. Its quantitative-claim detector extracts zero claims
from this paper's manuscript ($116\,202$ bytes, i.e.\ $116.2$\,kB, as
measured on 2026-09-22), yet reads the ``K'' in ``50K tokens'' as
kelvin. The two outcomes have the same cause: the detector reads only the
Abstract and Conclusions, and the paper's quantities do not appear
there in a form it recognises.

We repeated the detector test rather than rely on the earlier audit.
For the known-good control ``a band gap of 1.5\,eV,'' it returns one
claim, so the detector operates. It also returns one for ``50K
tokens.'' Yet it returns none for ``$r = 0.523$,'' ``1\% perplexity
loss,'' ``0.1 to 8B parameters,'' or ``$3\sigma$''---the vocabulary
used in this paper. A gate that runs, matches, and passes everything
differs from no gate only because users trust it. We therefore present
(v) as discipline maintained by an agent, not a mechanism, and treat
the undeployed gate as a result about mechanisation rather than an
unfinished feature.

Finally, they identify the file-drawer
problem as a limitation that pre-registration does \emph{not} solve
and offer registered reports as the remedy. Our append-only catalogue
is a complementary remedy: it keeps a failed mapping addressable even
when no venue publishes it. Their survey mentions effect-size
thresholds only once, as an evaluation issue orthogonal to
pre-registration. We discuss each addition below.

\paragraph{Pre-registration as git-timestamped YAML.} Each Phase-3
pre-registration is a plain-text YAML file committed to a public
repository \emph{before} the pilot. The ordering claim rests on the git
timestamp, which is a record rather than a gate. The pre-commit hook
written for this purpose (\texttt{prereg\_gate\_hook.py}) is registered
nowhere that this project loads, and we find no evidence that it ever
ran (\S\ref{sec:program}).

\paragraph{The $3\sigma$ effect-size gate.} A pilot PASSES only if
its headline effect exceeds baseline noise by $\geq 3\sigma$ across
$\geq 3$ random seeds, or across calibration bootstraps when seeds do
not apply. The gate acts on one pre-registered headline metric, not a
selected subset.

\paragraph{Append-only catalogue with negative-results-publishable
rule.} Every attempted mapping enters the catalogue, including
abandoned mappings, whose \texttt{lessons} fields must be populated. A
token-overlap novelty check on the \texttt{keywords} array detects
duplicate proposals. For each match, it returns the entry's status and
\texttt{lessons}, its recorded prior art, and, when the entry's own
novelty verdict was later back-tested, the outcome of that test. Thus,
an agent using the gate also learns whether it has failed before.

\emph{What the gate does not do.} The gate compares a proposal only
with \emph{recorded} prior art: our earlier attempts and published work
entered by an earlier agent. It does not search the literature. An
unrecorded pre-empting paper is therefore invisible, and a
\texttt{novel} verdict is not a finding of literature novelty. This
limitation had a cost. \citet{zagitov2026rethinking} pre-empts the
weight side of our rank-face result (Section~\ref{sec:related}), but
never entered the catalogue. We found it in a footnote to a later
survey \citep{tarasov2026tensor}, not through the gate. The Phase-1
survey is the remedy; the gate supplements rather than replaces it.
The caveat now also appears in the tool response, so even a caller who
reads only the verdict receives it.

\paragraph{Cluster-meta framing.} We place $N$ independent papers
under one falsifiable cluster prediction, here ``majority insulator,''
and use the joint outcome of the $N$ tests to evaluate it. The result is one $N$-test
ensemble experiment, not $N$ isolated papers, with $N$-fold power.

\paragraph{Cross-paper consistency checks.} When two papers measure
related physical quantities on one model, we pre-register a Pearson
correlation between them. A cluster-inconsistency event, such as the
cross-paper $r = -0.0583$ in Section~\ref{sec:cross_paper}, remains
informative even when the individual papers shelve.

\paragraph{Daemonic execution.} An autonomous research agent (Claude
Opus 4.7), running in a $24/7$ tmux session, generated, surveyed,
mapped, pre-registered, piloted, and adjudicated all five papers.
Standing agent discipline and git timestamps enforced pre-registration;
the hook written for that purpose is not wired (\S\ref{sec:program}).
The cluster-meta framing document
\fpath{projects/_cluster_meta/O_N_feasibility.md} was committed at
\texttt{defbb04} before any pilot. Human supervision was occasional
and limited to high-level direction and gating decisions. The agent
generated all experimental code and pre-registration content.

The Phase-3 compute budget also depended on one engineering change to
the Marzari--Vanderbilt $\Omega$ optimisation: precomputed
lag-covariance matrices gave a $\sim 200\times$ speed-up. Details
appear in Appendix~\ref{app:engineering}.

\section{Limitations}
\label{sec:limitations}

\paragraph{Scale: split by face.} The \emph{attention-regime}
inversion (P001, P002) was tested only on Pythia-160M and GPT-2-medium,
both with $\leq 350$M parameters. Llama-3-8B and larger models remain
untested on this face, leaving the $\geq 8$B-scale attention test open
(Section~\ref{sec:scale}). The \emph{rank-face} finding has broader
scale support. P011's embedding-rank negative replicated on OPT-1.3B,
and the failed MPO premise in P003-B held on Mistral-7B-v0.3 and
Qwen3-8B. Hence, tensor-network reshaping gives no compression over
plain SVD up to $7$--$8$B. This answers the scale question only in
part, and only for the rank face. Nor is its weight-side half a
priority claim: \citet{zagitov2026rethinking} established it earlier
and at larger scale (Section~\ref{sec:related}). The scopes must remain
separate. The follow-on studies in Section~\ref{sec:scale} measure
attention decay above $350$M, with P013 reaching $12$B, but only on
P002's decay observable. Because P001's Wannier-localisation arm has no
$\geq 8$B replication, the \emph{cluster} test of the attention face
still rests on the two $\leq 350$M checkpoints.

\paragraph{Positional-encoding artefacts.} Pythia-160M uses rotary
positional embedding (RoPE), whereas GPT-2-medium uses learned
absolute embedding; ALiBi was not tested \citep{press2022alibi}. A unitary
$U$ on the head-feature axis is incompatible with RoPE in the following sense:
RoPE acts on adjacent feature pairs, so an arbitrary unitary $U$ disrupts
positional encoding. This incompatibility
is a likely structural cause of the P001 failure. A future test should
either rotate in post-RoPE space by intercepting hidden states inside
the attention computation or constrain $U$ to block-diagonal forms
compatible with RoPE's 2D rotation blocks. The pre-registration
identified this risk in advance.

\paragraph{Specific calibration set.} All pilots used the
WikiText-103 test split to evaluate perplexity. We did not test domain
transfer to code, math, or long-context retrieval. The cluster-meta
framing document requires needle-in-haystack $\geq 64$k benchmarks
before any positive result ships, but the small models lack the
necessary context window. A future $\geq 8$B replication should add
these long-context tests.

\paragraph{Five physics frameworks, not exhaustive.} We tested
Wannier (P001), tight-binding (P002), DMRG (P003), Wilson RG (P005),
and tensor-train embeddings (P011). Physics-AI mappings also include
spin-glass replica theory, CFT, holographic RG, Anderson localisation,
and other frameworks. Our falsification concerns the insulator /
area-law analogy class only. It does not extend to those other anchors.

\paragraph{Pre-registration discipline depends on git-timestamp
trust.} The method assumes that repository history is not maliciously
rewritten between the pre-registration commit and the pilot. This is
uncontroversial for the present authors and audience. In an adversarial
setting, signed commits or external timestamping such as
OpenTimestamps would strengthen the protocol.

\section{Conclusion}
\label{sec:conclusion}

We tested a cluster of five solid-state-physics-inspired LLM
compression mappings on pretrained transformers. Four reached
pre-registered Phase-3 pilots; contemporaneous LLM-compression work
pre-empted one, P005, at Phase~1 by independently deriving the same
physics-anchored procedure without the physics rhetoric. All five
shelved. Data that we had committed to call SHELVE rather than reframe
falsified three Phase-3 pilots (P001, P002, P011). P003 shelved its
registered scaling claim but produced the first passing cross-paper
consistency check in the cluster.

Together, these outcomes invert the pre-registered cluster prediction
that most attention heads in pretrained small LMs behave like
Kohn-nearsighted insulators. Instead, they point toward critical,
glassy, or heavy-tailed regimes. The two cross-paper checks no longer sharpen that contrast: once padded positions are excluded from P002's histograms the insulator-observable pair gives $r = -0.0583$ and the critical-observable pair $r = 0.0472$, so both are null. Separately, on the weight/embedding-rank face,
tensor-network reshaping (P003-B MPO, P011 TT) gives no compression
over plain SVD from $0.1$ to $8$B. We observed this
``generic-tensor'' no-gain result three times. Its weight-side half was
established concurrently and independently by
\citet{zagitov2026rethinking}.

The central contribution is the method: git-timestamped YAML
pre-registration, a $3\sigma$ effect-size gate, an append-only
catalogue with mandatory \texttt{lessons} fields, cluster-meta framing
above individual papers, cross-paper consistency checks, and daemonic
execution under pre-registration discipline. It applies beyond these
five papers and beyond compression to any physics-AI mapping programme.
With this discipline, negative results are both publishable and
informative. Cluster-level inversions are more informative still.

Three recommendations follow. (i) A future compression-cluster
proposal must either test an \emph{attention}-regime claim at
$\geq 8$B in Phase 3, thereby testing the scale hypothesis of
Section~\ref{sec:scale}, or use critical / glassy / heavy-tailed
physics rather than insulator / area-law physics. Any
MPS/DMRG/area-law mapping must also identify genuine 1D locality in
the ML object, which the rank-face tests found in neither weights nor
embeddings. (ii) Before claiming novelty, a Phase-1 survey must compare
procedures side by side with SVD-LLM, AlphaPruning, OWL, and Redman et
al.\ (iii) The four pre-registered Phase-3 tests (P001, P002, P003,
P011) jointly falsify the cluster-meta majority-insulator prediction.
Their data trail---results JSONs, pre-registration files, verdict
files, and catalogue entries---is recorded at the commit hashes cited
throughout the paper and in
\fpath{projects/M01_methodology_paper/drafts/provenance.md}.

\section*{Data and code availability}

The \fpath{claude-airesearch} repository contains all experimental
code, raw results, pilot figures, pre-registration files, and
catalogue entries. The relevant paths are
\fpath{projects/{P001,P002,P003,P005,P011}/},
\fpath{physics_redundancy_catalog.json}, and
\fpath{pre_registered_hypotheses/}. The provenance map from every
numerical claim to its source artefact is
\fpath{projects/M01_methodology_paper/drafts/provenance.md}. The
pre-registration commits are \texttt{10f2330} (P001, P002),
\texttt{0e37eba} (P003), and \texttt{97911fb} (P011). The P002 SHELVE
commit is \texttt{aeda9e4}, the P001 SHELVE commit is
\texttt{f0da3c0}, the P011 SHELVE commit is \texttt{b7a0cf2}, and the
P003 stage-A adjudication is \texttt{0614b5c}.

\section*{Acknowledgments}

The author gratefully acknowledges the steadfast, long-term support of
the Department of Physics, the Faculty of Arts and Sciences, and the
University of Puerto Rico at Mayag\"uez.

The \texttt{claude-airesearch} agent programme ran the reported
experiments. Computation was local on a Mac Studio M3 Ultra (Apple
Silicon, MPS backend). Across the four Phase-3 pilots, catalogued
wall-clock was $0.299$ device-hours for P001 Wannier on Pythia-160M,
$0.128$ device-hours for P002 tight-binding on GPT-2-medium and
Pythia-160M, $3.4$ device-hours for P003 DMRG stage-A on GPT-2-medium,
Qwen3-8B and Mistral-7B-v0.3, and $0.6$ device-hours for P011
tensor-train embedding on GPT-2 and OPT-1.3B. The total was $4.427$
MPS-device-hours, dominated by P003 stage-A, plus Phase-1 survey time
of $\sim$1--3 agent-hours per paper, which was not GPU-bound. These
figures are wall-clock measurements from one M3 Ultra $+$ MPS device
and are not directly comparable with NVIDIA A100/H100 GPU-hour
accounting. The author thanks the developers of the open-source probe
models---Pythia by EleutherAI, GPT-2 by OpenAI, OPT by Meta AI, Qwen by
Alibaba, and Mistral by Mistral AI---and the maintainers of the
\textsc{WikiText-103} corpus.

\bibliographystyle{plainnat}
\bibliography{refs}

\appendix

\section{Engineering note: precomputed lag-covariance matrices for the
Marzari--Vanderbilt $\Omega$ functional}
\label{app:engineering}

Our first implementation of the Marzari--Vanderbilt $\Omega$
functional recomputed the FFT of the rotated activations at every Adam
step. This gave a $\sim 6$-hour fit time per head on Pythia-160M. We
replaced the per-step FFT with precomputed lag-covariance matrices,
\[
  M^{(h),\ell}(\tau) = \mathbb{E}_t\!\left[a^{(h),\ell}(t)\,a^{(h),\ell}(t + \tau)^\top\right]
\]
and used the closed form
\[
  \Omega(U) = \sum_\alpha \big[U^\top M(\tau)\,U\big]_{\alpha\alpha}\cdot \tau^2 \,/\, \text{norm}
\]
This reduces the per-head fit from $\sim 150$ seconds to
$\sim 0.75$ seconds, validated to $10^{-5}$ against direct activation
rotation, for a $\sim 200\times$ speedup. The Wannier optimisation for
the full pilot fell from $\sim 6$ hours to $\sim 7$ minutes. This is a
footnote-grade engineering contribution: it kept the Phase-3 pilot
within a single afternoon's compute budget but is incidental to the
paper's methodological claims.

\end{document}